\documentclass[aps,prd,twocolumn,superscriptaddress,nofootinbib,longbibliography]{revtex4-2}
\pdfoutput=1

\usepackage{amsmath,amssymb}
\usepackage{amsthm}
\usepackage[utf8]{inputenc}
\usepackage[english]{babel}
\usepackage{graphicx}
\usepackage{dcolumn}

\usepackage{ulem}
\usepackage{bm}
\usepackage{braket}
\usepackage{mathtools}
\usepackage{url}
\usepackage{soul}
\usepackage{array}
\usepackage[colorlinks,citecolor=blue,linkcolor=blue,urlcolor=blue, breaklinks=true]{hyperref}
\usepackage{float}
\usepackage{orcidlink}
\makeatletter
\AtBeginDocument{\let\LS@rot\@undefined}
\makeatother

\usepackage{comment}

\usepackage{xcolor}
\usepackage{xspace}
\usepackage{ifthen}
\graphicspath{{/}}

\newcommand{\showcomments}{true}

\newcommand{\marios}[1]%
{\ifthenelse{\equal{\showcomments}{true}}%
{{\color{blue}{\small \textbf{M:} #1}}}{\xspace}}%

\newcommand{\unknown}[1]%
{\ifthenelse{\equal{\showcomments}{true}}%
{{\color{orange}{\small \textbf{?:} #1}}}{\xspace}}%

\usepackage{physics}
\newcommand{\mf}{\mathsf}
\newcommand{\tc}[1]{\textsc{#1}}

\begin{document}

\author{Nikolaos Mitrakos}
\affiliation{Institute for Quantum Optics and Quantum Information (IQOQI) Vienna, Austrian Academy of Sciences, Boltzmanngasse 3, A-1090 Vienna, Austria}

\author{Maria Papageorgiou}
\affiliation{Institute for Quantum Optics and Quantum Information (IQOQI) Vienna, Austrian Academy of Sciences, Boltzmanngasse 3, A-1090 Vienna, Austria}

\author{T. Rick Perche}
\affiliation{Nordita,
KTH Royal Institute of Technology and Stockholm University,
Hannes Alfvéns väg 12, 23, SE-106 91 Stockholm, Sweden}

\author{Marios Christodoulou}

\affiliation{Institute for Quantum Optics and Quantum Information (IQOQI) Vienna, Austrian Academy of Sciences, Boltzmanngasse 3, A-1090 Vienna, Austria}

\begin{abstract} \noindent

Detection of entanglement through the Newtonian potential has been claimed to support the existence of gravitons, by extrapolating to a thought experiment which demonstrates that complementarity and causality would be in conflict unless quantum fluctuations exist. We critically assess this consistency argument using scalar field models. 
 We show that whether complementarity or no-signalling is violated when quantum fluctuations are neglected, depends on how this approximation is taken, while in both cases entanglement is generated locally in spacetime. We clarify that the correct reading of the paradox requires making a clear distinction between two notions of causality violation:
Newtonian action-at-a-distance and quantum mechanical signalling at spacelike separation; the latter is pertinent while the former is not. We conclude that the thought experiment (a) does not add to the epistemological relevance of entanglement through Newtonian potentials (b) lends support for the existence of gravitons, if retardation effects are detected in entanglement through gravity. 
\end{abstract}

\date{\today}

\title{When does entanglement through gravity imply gravitons?}
\maketitle 
\bigskip

\section{Introduction}

Detecting entanglement generation through the gravitational interaction is believed to provide some evidence for quantum aspects of the gravitational interaction e.g.~\cite{Marletto:2024ltk,Bose:2024nhv}. Intuitively, for the effect to arise, the gravitational field gets entangled with quantum matter, which implies it can exist in a quantum superposition \cite{Christodoulou:2018cmk,Bengyat:2023hxs}. This can not happen in a classical field theory like general relativity, thereby falsifying the theory---a result of extraordinary importance. Whether this conclusion is compelling remains debatable.

A line of argument that has received attention employs a consistency argument presented through a thought experiment, which involves entanglement-through-gravity of two distant parties. It exemplifies an apparent tension between causality and complementarity, resolved when quantum fluctuations are taken into account \cite{2016mariExperimentsTestingMacroscopic,2018belenchiaQuantumSuperpositionMassive, 2019belenchiaInformationContentGravitational, 2022danielsonGravitationallyMediatedEntanglement}. Then, the argument goes,
if entanglement-through-gravity is detected, it must be that quantum field excitations---gravitons---exist.

In this work, we critically assess the relevance of this thought experiment for the epistemological importance of detecting entanglement-through-gravity using scalar field models. We build on the demonstration that in a relativistic setting entanglement in table-top quantum gravity protocols is generated with retardation according to the light-cone structure \cite{Christodoulou:2022mkf}, and the in-depth examination of quantum field–theoretic models applicable to the thought experiment carried out in \cite{eirini, 2022hidaka,Hidaka:2022gsv,Uncertainty1, Uncertainty2}. 

Our main result (Sections \ref{sec:retrocausality} and \ref{sec:Paradox}) is to demonstrate that, when quantum fluctuations are neglected in a quantum field theory, whether complementarity or no-signalling is violated depends on how this approximation is implemented. We distinguish two cases (i) forcing the field anti-commutator (the Hadamard function) to be zero, which breaks complementarity while no-signalling holds (ii) taking a stationary phase approximation, which implies that both the Hadamard function and the so-called causal propagator are neglected, which allows for signalling at spacelike separation while complementarity holds. In both approximations entanglement is generated locally in spacetime, through phases that correspond to retarded propagators. 

We clarify misconceptions in readings of this paradox that stem from assuming Newtonian action at a distance, i.e.~instantaneous generation of entanglement (Sections \ref{sec:NoPertinenceNewton} and \ref{sec:NoSpacelikeInfluences}). We conclude that if \textit{retardation} in the generation of entanglement is detected (protocols to this effect have been discussed in \cite{Christodoulou:2022mkf,Bengyat:2023hxs}), such experimental data, in conjunction with the consistency argument arising from the thought experiment, would support the existence of gravitons.

The discussion will focus on the form of the final reduced density matrices of the two parties. To introduce gradually the formalism, we first treat a toy-model (Sections \ref{sec:Scalar} and \ref{sec:propagators}), and then discuss how the analysis extends to a more realistic description (Section \ref{sec:pathModel}).



\section{Toy model for entanglement through a scalar field}
\label{sec:Scalar}

The main features of quantum field theory needed for the discussion are now introduced using a toy-model description of entanglement through a quantum field, proposed and studied in detail in \cite{2023hidaka}. The original derivation used here is given in Appendix \ref{appendix: Magnus}, only main steps are sketched below.

Consider two masses A and B, and a scalar quantum field. The masses are static, their positions are fixed once and for all at $\boldsymbol{x}_{\tc a}$ and $\boldsymbol{x}_{\tc b}$, and they possess `internal' spin-$1/2$ degrees of freedom $s_\textsc{a}$ and $s_\textsc{b}$. The combined system of the two `spins with mass' and the quantum field $\phi$, lives in a Hilbert space $
\mathcal{H}=\mathbb{C}^2\otimes\mathbb{C}^2\otimes \mathcal{F}_{{\phi}}$, with $\mathcal{F}_{{\phi}}$ the Fock space of $\phi$. Initially, the masses and field are in a product state, and the spins are assumed in superposition
\begin{equation}
\label{eq:instate}
\ket{\Psi(t_i)}=\frac{1}{2}\sum_\sigma  \ket{\sigma}\otimes \ket{\omega},
\end{equation}
where $\ket{\omega}$ is the vacuum state of the scalar field, and $\ket{\sigma}=\ket{s_\textsc{a} s_\textsc{b}}$ are the spin configurations, with $\sigma =\{s_{\tc a} s_{\tc b}\}\in \{00,10,01,11\}$. 

The interaction Hamiltonian (in the interaction picture) is 
\begin{equation}
\label{eq:artificialHI}
     \hat{{H}}_{int}(t) = -\lambda_{\tc{a}}(t)\hat{\sigma}^z_{\tc{a}}\hat{\phi}(\mf x_\tc{a})-\lambda_{\tc{b}}(t)\hat{\sigma}^z_{\tc{b}}\hat{\phi}(\mf x_\tc{b}).
\end{equation}
where $\hat{\sigma}^z_a=\ket{0}\!\!\bra{0}_a-\ket{1}\!\!\bra{1}_a$ is the third Pauli matrix acting on the spin $a=\text{A},\text{B}$ and $\mf x_{a} = (t,\bm x_a)$. The `switching' functions $\lambda_{a}(t)$ encapsulate both the coupling strength and the duration of the interaction. They are non-zero only in a time interval (which can be different for each mass), effectively `switching on' the mass-field interaction during that time.

The physics of \eqref{eq:artificialHI} are made more transparent by viewing it as the spatial integral of the Hamiltonian density
\begin{equation}\label{eq:fullartificialHI}
 \hat{{h}}_{int,a}(\mf x)=-\sum_{s_a=0}^1 \lambda_{a,s_a}(t) \delta^3(\boldsymbol{x}-\boldsymbol{x}_{s_a}^{a}(t))\, \ket{s_a}\!\!\bra{s_a}\otimes \hat{\phi}(\mf x)
\end{equation}
where $\boldsymbol{x}_{s_a}^{a}(t)$ is the classical trajectory that each mass follows for a specific value of its spin state $s_a\in\left\{0,1\right\}$.\footnote{It is implied that the coupling between mass A and the field acts trivially on mass B, $\ket{s_{\textsc{a}}}\bra{s_{\textsc{a}}}\otimes\openone_{\textsc{b}}\otimes\hat{\phi}({\mf x})$, and vice versa.} Then, \eqref{eq:artificialHI} corresponds to setting that: (a) $\bm x^{a}_{0}(t)$ = $\bm x^{a}_{1}(t) = \bm x_{a}$ in \eqref{eq:fullartificialHI}, i.e.~the field is coupled \textit{locally} to each mass, and (b) the couplings for $s_a=0$ and $s_a=1$ only differ by a sign  $\lambda_{a,1}(t)=-\lambda_{a,0}(t)\equiv \lambda_a(t)$.  Since the spin states are initially in superposition, see \eqref{eq:instate}, evolution yields a superposition of coherent states (which will only differ by overall signs of the displacement operators).

 In the interaction picture, the evolution of the total state of the system is given by $\ket{\Psi(t_{f})}=\hat{U}(t_{f},t_{i}) \ket{\Psi(t_{i})}$, where $\hat{U}(t_{f},t_{i})$ is the time-ordered exponential of \eqref{eq:artificialHI}.  
 The evolution operator takes the form
 \begin{equation}
\label{eq:parApprox}
\hat{U}(t_{f},t_{i})=\sum_{\sigma}\ket{\sigma}\!\!\bra{\sigma}\otimes \hat{U}_{\phi}^\sigma (t_{f},t_{i}),
\end{equation}
 with the   `per-branch' field evolution operators 
\begin{equation} \label{time-ordered}
 \hat{U}_{\phi}^\sigma (t_{f},t_{i})   = \mathcal{T}e^{-i\int^{t_f}_{t_i} \sum_a s_a\lambda_a(t) \hat{\phi}(t,\bm{x}_a) \dd t}.
\end{equation}

To find the final state of the two masses, we trace out the field to get
\begin{equation}
\label{eq:toyrhoevolution}
\hat{\rho}^f_{\tc{ab}}=\frac{1}{4}\sum_{\sigma, \sigma'} \langle \omega^f_{\sigma'}|\omega^f_{\sigma}\rangle   \; \ket{\sigma}\bra{\sigma'}.  
\end{equation}
with $|\omega^f_{\sigma}\rangle:=\hat{U}^{\sigma}_\phi (t_f,t_i)\ket{\omega}$ the evolved field vacuum, a coherent state that depends on $\sigma$.\footnote{Using \eqref{time-ordered} and the standard mode decomposition $\hat{\phi}(\mathsf x)=\int \dd \bm{k} \,u_{\bm{k}} (\mathsf x) \hat{a}_{\bm{k}}+$ h.c., $|\omega^f_{\sigma}\rangle= \hat{D}[\lambda]\ket{\omega}$, with $\hat{D}[\lambda]= e^{\hat{a}^{\dagger}[\lambda]-\hat{a}[\lambda]}$ and $\hat{a}[\lambda]=\int \dd \bm{k} \lambda(\bm{k}) \hat{a}_{\bm{k}}$, with $\lambda(\bm{k})=i \sum_a s_a \int \dd t \lambda_a(t) u^*_{\bm{k}}(t, \bm{x}_a)$.}
The (non perturbative) calculation of the overlaps 
\begin{equation} \label{overlaps}
\langle \omega^f_{\sigma'} |\omega^f_{\sigma}\rangle=\langle \omega| (\hat{U}^{\sigma'}_\phi (t_f,t_i))^{\dagger}\hat{U}^{\sigma}_\phi (t_f,t_i)|\omega\rangle.
\end{equation}
 is done through the Magnus expansion, a technique convenient for dealing with the time-ordered unitaries $\hat{U}^{\sigma}_\phi (t_f,t_i)$.
 
The evolved joint state of the two masses is
\begin{equation}
\label{eq:toyrhofFull}
\begin{matrix}
4 \hat{\rho}^f_\tc{ab} = \left( \begin{matrix}
1\quad\quad & e^{-\gamma_{\tc{b}}}\; e^{ 2i \varphi^{\tc b}_{\tc a} }      &  e^{-\gamma_{\tc{a}}} \; e^{ 2i \varphi^{\tc a}_{\tc b}    }   & \quad e^{-\left(\gamma_{\tc{a}}+\gamma_{\tc{b}}+2\gamma_{\tc{ab}}\right)}   \\
&  1 & \quad e^{-\left(\gamma_{\tc{a}}+\gamma_{\tc{b}}-2\gamma_{\tc{ab}}\right)}  &  e^{-\gamma_{\tc{a}}} \; e^{ -2i \varphi^{\tc a}_{\tc b}    }  \\
&   &   1  &e^{-\gamma_{\tc{b}}}\; e^{ -2i \varphi^{\tc b}_{\tc a} } \\
  &   &  &  1
\end{matrix} \right) 
\end{matrix} \\
\end{equation}
where we defined 
\begin{align}
\label{eq:toyGHterms}
\varphi^{\tc a}_{\tc b}& =\iint \dd t\,\dd t'\;\lambda_{\tc a}(t)G_{\tc r}(t,\boldsymbol{x}_{\tc a}\,;t',\bm x_{\tc b})\,\lambda_{\tc b}(t')\nonumber\\
\varphi^{\tc b}_{\tc a}& =\iint \dd t\,\dd t'\;\lambda_{\tc b}(t)G_{\tc r}(t,\boldsymbol{x}_{\tc b}\,;t',\bm x_{\tc a})\,\lambda_{\tc a}(t')\nonumber\\
\gamma_{\tc a \tc b}& =\iint \dd t\,\dd t'\;\lambda_\tc{a}(t)H(t,\boldsymbol{x}_a\,;t',\bm x_b)\,\lambda_\tc{b}(t') \nonumber \\
\gamma_{a}& =\iint \dd t\,\dd t'\;\lambda_a(t)H(t,\boldsymbol{x}_a\,;t',\bm x_a)\,\lambda_a(t').
\end{align}
for $a=\text{A,B}$. Here, $G_{\tc r}(t, \boldsymbol{x}_{\tc b}\,;t', \bm x_{\tc a})$ is the field's retarded propagator, and $H(t, \boldsymbol{x}_{\tc b}\,;t', \bm x_{\tc a})$ the field's Hadamard function, evaluated on the locations of the two masses. The roles of $G_{\tc r}$ and $H$ are discussed in a moment, in Section \ref{sec:propagators}. We have introduced the convention that lower indices on phases $\varphi$ denote \textit{source} and upper indices denote \textit{recipient}.

The reduced density matrix for A when tracing out B reads
\begin{equation}
\label{eq:reducedAtoy}
\hat{\rho}_\tc{a}(t_f)= 
\frac{1}{2}\begin{pmatrix}
1  & e^{-\gamma_{\tc{a}}}  \cos{ (2 \varphi^{\tc a}_{\tc b} )   }\\ 
 e^{-\gamma_{\tc{a}}}  \cos{ (2 \varphi^{\tc a}_{\tc b}  )  } & 1
\end{pmatrix} 
\end{equation}
Similarly, the reduced density matrix for mass B is given by $A \leftrightarrow B$.  We see that: (a) the pure phases $\varphi^{\tc a}_{\tc b}$ will create \textit{entanglement} between A and B, (b) the exponents $\gamma_{\tc{a}}$, $\gamma_{\tc{b}}$ cause \textit{local decoherence}, due to the local to A or B coupling to the field, (3) the exponents $\gamma_{\tc a \tc b}$ are not locally accessible to  A and B. 

\section{Causal Influences and Quantum Fluctuations}
\label{sec:propagators}
We now discuss the meaning of the phases $\varphi^\tc{a}$, $\varphi^\tc{b}$ and the exponentially suppressing exponents $\gamma_{\tc{a}}$, $\gamma_{\tc{b}}$ and $\gamma_{\tc{ab}}$, that appear in the final state \eqref{eq:toyrhofFull}.

{\bf Retarded propagation of which path information.} The retarded propagator $G_{\tc{r}}(\mf x;\mf x')=G_{\tc{r}}(t, \bm{x}; t', \bm{x}')$ is given by the time-ordered commutator $G_{\tc{r}}(\mf x;\mf x')=-i \langle[\hat{\phi}(\mf x), \hat{\phi}(\mf x') ]\rangle\theta(t-t')$. It describes causal propagation from the spacetime point $\mf x'$ to a point $\mf x$ inside the future lightcone of $\mf x'$. It solves the \textit{classical} equations of motion: it does \textit{not} encode quantum fluctuations. 

For example, the phase $\varphi^{\tc a}_{\tc b}$, see \eqref{eq:toyGHterms}, describes the causal influence of  B on A. The interaction of B with the field is encoded in the switching function $\lambda_{\tc b}(t')$. The convolution of $\lambda_{\tc b}(t')$ with the retarded propagator sourced by B is supported in the causal future of B. If A intersects the causal future of B, their joint state \eqref{eq:toyrhofFull}, and the reduced state of A \eqref{eq:reducedAtoy}, accumulate a phase $\varphi^{\tc a}_{\tc b}$. These phases create the entanglement-through-gravity between A and B discussed in table-top quantum gravity. It is generated locally in spacetime, with retardation \cite{Christodoulou:2022mkf,GMEEdu2023}.

A different way to see that the retarded and advanced propagators do not encode quantum fluctuations is through
\begin{equation} \label{commutator}
    [\hat{\phi}(\mathsf x), \hat{\phi}(\mathsf x')]= i  E(\mathsf x ; \mathsf x') \openone
\end{equation}
where $E(\mathsf x ; \mathsf x')= G_{\tc{r}} (\mathsf x; \mathsf x ') - G_{\tc{a}} (\mathsf x; \mathsf x ')$ and $\openone$ the identity operator on the field. Here, $G_{\tc{a}}$ is the advanced propagator and $E$ is known as the causal propagator. The causal, retarded and advanced propagators are numbers; they are trivial operators on the state of the quantum field. Although $E$ and $G_{\tc{a}}$ do not appear in the final state \eqref{eq:toyrhofFull}, they are needed for deriving it, they play a role in the evolution. We will see in Section \ref{sec:retrocausality} that the stationary phase approximation, in addition to neglecting the Hadamard function $H$, also forces $E$ to zero. 

\medskip
{\bf Quantum fluctuations.} The Hadamard function is the expectation value of the field anti--commutator, $H(\mf x;\mf x')=\langle{\{\hat{\phi}(\mf x),\hat{\phi}(\mf x')\}}\rangle$. Unlike the commutator, the anti--commutator is a non--trivial functional of the quantum field: it is not proportional to the identity operator. It encodes the effects of quantum fluctuations. 

The exponents $\gamma_{\textsc{a}}$, $\gamma_{\textsc{b}}$ and $ \gamma_{\tc{ab}}$, describe (scalar) quantised radiation production~\cite{2001breuer,2013blencowe,2017bassi, 2021kanno,2024hsiang}. When only A or B is coupled to the field, the expected number of produced (scalar) particles is $\langle\hat{N}\rangle = \gamma_\tc{a}$ or $\langle\hat{N}\rangle = \gamma_\tc{b}$, respectively, where $\hat{N}$ is the total particle number operator for the scalar field \cite{2023hidaka}. When both masses are coupled to the field, we simply have $\langle\hat{N}\rangle = \gamma_\tc{a}+\gamma_\tc{b} $ (the $\gamma_\tc{ab}$ do not contribute).

The exponents $\gamma_\tc{a}$ and $\gamma_\tc{b}$ cause decoherence local to A or B, due to radiation that arises from switching `on and off' the local coupling to the quantum field. The exponent $\gamma_{\textsc{ab}}=\gamma_{\textsc{ba}}$ encodes correlations between the masses that are present even when the masses are spacelike separated, since the Hadamard function does not vanish when evaluated at spacelike separated points \cite{2024percheb}. They correspond to correlations already present between pairs of points in the vacuum state, which get `picked up' by the masses through their coupling to the field, see e.g.~\cite{2015pozas-kerstjens}. There are scenarios in which these correlations lead to entanglement between the masses. This phenomenon, known as `entanglement harvesting' \cite{2015pozas-kerstjens,Pozas2016,SachsMannEdu2017,2021tjoa,hectorHarvesting2022,borisHarvesting2023,2024perchea}, is far removed from the regime of table-top entanglement-through-gravity. Since $\gamma_{\textsc{ab}}$ only appears on the anti-diagonal of the final state \eqref{eq:toyrhofFull}, it and drops out from the reduced state \eqref{eq:reducedAtoy}. Therefore, these correlations are not accessible locally by A or B. The decoherence relevant for the resolution of the paradox discussed in Section \ref{sec:Paradox} is caused by $\gamma_\tc{a}$ and $\gamma_\tc{b}$; $\gamma_{\tc{ab}}$ play no role.

\section{Including path superposition}\label{sec:pathModel}
The toy-model of the preceding Sections can be generalised so that each mass is in a path superposition. Similar models have been studied in \cite{SugiyamaFirst,Uncertainty1, Uncertainty2, Uncertainty3}. The derivation used here is given in Appendix \ref{appendix: Magnus}. Consider two masses, A and B, each with a spin-$1/2$ degree of freedom. Assuming a superposition of spin states at initial time $t_i$,
\begin{equation}
\label{eq:II_spins}
\ket{s(t_i)}_{\tc a}=\ket{s(t_i)}_{\tc b}= \frac{1}{\sqrt{ 2 }}\left( \ket{0}+\ket{1}   \right) 
\end{equation}
and imagining this state traversing a region with an inhomogeneous external magnetic field, each mass will follow a superposition of paths. Without specifying details, we model this formally by a superposition of arbitrary classical trajectories $\boldsymbol{x}^{a}_{s}(t)$, $a=$ A, B, one for each spin value $s=0$ and $s=1$. The trajectory of each mass is determined by its spin, it is in the left position $\ket{L}$ when its spin is $\ket{0}$ and in the right position $\ket{R}$ for spin $\ket{1}$. For brevity, we identify the spin degree of freedom with the corresponding path and write $s = \text{L},\text{R}$.

 The particles couple to the field through the interaction Hamiltonians
\begin{align}
    \hat{H}_\tc{a}(t) = \lambda_\tc{a} \left( \ket{L_\tc{a}}\!\!\bra{L_\tc{a}} \hat{\phi}(t,\bm x_{\tc l}^{\tc{a}}(t))+ \ket{R_\tc{a}}\!\!\bra{R_\tc{a}} \hat{\phi}(t,\bm x_{\tc r}^{\tc{a}}(t)) \right)\nonumber\\
    \hat{H}_\tc{b}(t) = \lambda_\tc{b} \left( \ket{L_\tc{b}}\!\!\bra{L_\tc{b}} \hat{\phi}(t,\bm x_{\tc l}^{\tc{b}}(t))+ \ket{R_\tc{b}}\!\!\bra{R_\tc{b}} \hat{\phi}(t,\bm x_{\tc r}^{\tc{b}}(t)) \right).  \label{eq:HAHB}
\end{align}
The relevant period of the interaction is in a time interval $t \in \{t_r, t_f\}$, when the trajectories $\bm x^a_{\tc l}(t)$ and $\bm x_{\tc r}^a(t)$ are assumed distinct. The interaction Hamiltonians $\hat{H}_a(t)$ trivialize to $\lambda_{a}\openone_{a}\hat{\phi}(t,\bm x_\tc{l}^{a}(t))$ when $\bm x^a_\tc{l}(t) = \bm x_\tc{r}^a(t)$.

The derivation proceeds similarly to Section \ref{sec:Scalar}, the resulting expressions are more involved because each mass couples to the field along two different trajectories. The state of the two masses at the final time $t_f$ is 
\begin{equation}
4 \hat{\rho}^f_\tc{ab}= \left( \begin{matrix}
\; 1& \quad e^{-\Gamma_{\tc{b} }}\, e^{ i \varphi^\tc{b}_{\tc{a}\tc{l}} }      & \quad e^{-\Gamma_{\tc{a}}} \, e^{ i \varphi^{\tc a}_{\tc{bl}}    }   & \quad Y    \\
&  1  & X  & \quad e^{-\Gamma_{\tc{a}}} \, e^{ i \varphi^{\tc a}_{\tc b\tc{r}}    }  \\
&   &   1  & \quad e^{-\Gamma_{\tc{b}}}\, e^{ i \varphi^{\tc b}_{\tc{ar}} } \\
  &   &  &  \quad 1 \\ 
\end{matrix} \right) 
\label{eq:rhofpath}
\end{equation}
with
\begin{align}
  X&= \exp(-\left(\Gamma_{\tc{a}}+\Gamma_{\tc{b}}-2\Gamma_{\tc{ab}}\right) + \frac{i}{2} \sum_{s}(\varphi^{\tc{a}}_{\tc{b}s}-\varphi^{\tc b}_{\tc{a}s} ) )
\nonumber \\
  Y&= \exp(-\left(\Gamma_{\tc{a}}+\Gamma_{\tc{b}}+2\Gamma_{\tc{ab}}\right) + \frac{i}{2} \sum_{s}(\varphi^{\tc{a}}_{\tc{b}s}+\varphi^{\tc b}_{\tc{a}s} ) )
\end{align}
where we have defined
\begin{equation}
    \begin{aligned}
    \label{eq:obsPhases}
\varphi^{\tc{a}}_{\tc{b}s}&=\varphi^{\tc{al}}_{\tc{b}s}-\varphi^{\tc{ar}}_{\tc{b}s} \\
    \varphi^{\tc{b}}_{\tc{a}s}&=\varphi^{\tc{bl}}_{\tc{a}s}-\varphi^{\tc{br}}_{\tc{a}s}
    \end{aligned}
\end{equation}
with $s_a\in\{\tc{l},\tc{r}\}$ for $a=\text{A},\text{B}$,
and
\begin{equation}
\begin{aligned}
 \label{eq:bigGamma}   4 \Gamma_{\tc{a}}&=\gamma_{\tc{a}}^{\tc{ll}} +\gamma_{\tc{a}}^{\tc{rr}} -2 \gamma_{\tc{a}}^{\tc{lr}} \\
    4\Gamma_{\tc{b}}&=\gamma_{\tc{b}}^{\tc{ll}} +\gamma_{\tc{b}}^{\tc{rr}} -2 \gamma_{\tc{b}}^{\tc{lr}} \\
    4\Gamma_{\tc{ab}}&=\gamma_{\tc{ab}}^{\tc{ll}} +\gamma_{\tc{ab}}^{\tc{rr}} - \gamma_{\tc{ab}}^{\tc{lr}} - \gamma_{\tc{ab}}^{\tc{rl}}.
\end{aligned}
\end{equation}

The structure is similar to the final state \eqref{eq:toyrhofFull} for the toy model of Section \ref{sec:Scalar}, and the morals discussed in Section \ref{sec:propagators} carry through. There are now four retarded propagation phases $\varphi^{as_a}_{b s_b}$, one per spin branch $\sigma =\{s_{\tc a} s_{\tc b}\}\in \{00,10,01,11\}$ when A is the source and another four when B is the source.
Recall that a lower subscript $b=\text{A}, \text{B}$ denotes the source of the causal influence on $a=\text{B}, \text{A}$ respectively. For example, if A is in the left path $\ket{L}$ and mass B is in the right path $\ket{R}$, we get the phase $\varphi^{\tc{br}}_{\tc{al}}$ given by
\begin{equation} \label{phiLR}
    \varphi^{\tc{br}}_{\tc{al}}=\iint \dd t\,\dd t'\;\lambda_\tc{b} G_{\tc r}(t, \bm x_{\tc r}^{\tc{b}}(t)\,;t', \bm x_{\tc l}^{\tc{a}} (t'))\lambda_\tc{a}.
\end{equation} 
If both masses were in a definite path this would be a global phase. Since only \textit{relative} phases are observable, the eight phases $\varphi^{as_a}_{b s_b}$ appear in equation~\eqref{eq:rhofpath} only in the four linearly independent combinations defined in \eqref{eq:obsPhases}; these are the phase differences that are observable. For example, $\varphi^{\tc{b}}_{\tc{al}} = \varphi^{\tc{bl}}_{\tc{al}} -\varphi^{\tc{br}}_{\tc{al}}$ is the measurable causal influence that mass A in path $\ket{L}$ has on mass B.

The eight exponents $\gamma^\sigma_a$ are integrals of the Hadamard function. Only six are distinct because $\gamma_{\tc{a}}^{\tc{lr}}=\gamma_{\tc{a}}^{\tc{rl}}$. They appear in the final state only in the two combinations $\Gamma_\tc{a}$ and $\Gamma_\tc{b}$ defined in  \eqref{eq:bigGamma}. $\Gamma_\tc{a}$ and $\Gamma_\tc{b}$ encode decoherence due to quantised radiation arising from the spacetime local coupling of each mass to the field. This is shown in Appendix~\ref{sec:AppFluct}. The exponents $\Gamma_\tc{ab}$ include spacelike correlations present in the vacuum state of the field and obtained by the masses, but will not appear in the reduced density matrices of each mass. The $\Gamma_\tc{ab}$ play no role in the thought experiment discussed in what follows.

The above remarks can be confirmed by inspection of the reduced density matrix for A, which reads
\begin{equation}
\label{eq:reducedA}
\hat{\rho}_\tc{a}(t_f)= 
\frac{1}{2}\begin{pmatrix}
1  & \frac{1}{2} e^{-\Gamma_{\tc{a}}}  (e^{ i \varphi^{\tc a}_{\tc{bl}}}+e^{ i \varphi^{\tc a}_{\tc{br}}})\\ 
\frac{1}{2} e^{-\Gamma_{\tc{a}}}  (e^{ -i \varphi^{\tc a}_{\tc{bl}}}+e^{- i \varphi^{\tc a}_{\tc{br}}}) & 1
\end{pmatrix}, 
\end{equation}
compare with \eqref{eq:reducedAtoy}. It includes quantum fluctuations created locally at A, encoded by $\Gamma_\tc{a}$, and causal influences from B encoded in $\varphi^{\tc{a}}_{\tc{b}s_{\textsc{b}}}$. An analogous discussion applies to the reduced density matrix of B, upon swapping  $ A \leftrightarrow B$.

\section{What breaks when neglecting quantum fluctuations}\label{sec:retrocausality}
 We now ask what fails when quantum fluctuations are neglected. There are two obvious implementations of this approximation: (i) setting them to zero directly, and (ii) taking a stationary-phase approximation. These give rise to distinct problems.

{\bf (i).~Negative probabilities when forcing coherence ad-hoc.} 
Quantum fluctuations can be neglected by setting $\Gamma_\tc{a}=\Gamma_\tc{b}=0$ in the final state \eqref{eq:rhofpath}~\cite{theRole2023}, which implies also $\Gamma_{\tc{ab}}=0$ since $\Gamma_{\tc a \tc b}\leq \frac{1}{2}(\Gamma_{\tc a }+\Gamma_{\tc b })$, see Appendix~\ref{sec:AppFluct}. It corresponds to forcing to zero the Hadamard function $H(\mf x;\mf x') = \langle{\{\hat{\phi}(\mf x),\hat{\phi}(\mf x')\}\rangle}$, see \eqref{eq:toyGHterms}, \eqref{eq:bigGamma} and \eqref{eq:HadamardDef}. The final state \eqref{eq:rhofpath} becomes
\begin{equation}
\begin{matrix}
4\hat{\rho}^f_\tc{ab}=\left( \begin{matrix}
1\quad  &  e^{ i \varphi^{\tc b}_{\tc{al}} }      &  e^{ i \varphi^{\tc a}_{\tc{bl}}    }   &  e^{\frac{i}{2} \sum_{\sigma}(\varphi^{\tc{ a}}_{\tc{b}\sigma}+\varphi^{\tc b}_{\tc{a}\sigma} )}  \\
&  1  &  e^{\frac{i}{2} \sum_{\sigma}(\varphi^{\tc{ a}}_{\tc{b}\sigma}-\varphi^{\tc b}_{\tc{a}\sigma} )} &  e^{ i \varphi^{\tc a}_{\tc{br}}    }  \\
&   &   1  &  e^{ i \varphi^{\tc b}_{\tc{ar}} } \\
  &   &  &  1
\end{matrix} \right) 
\end{matrix}.\label{eq:rhofnogamma}
\end{equation}
 
Inspection of the above density matrix, immediately suggests something is off.  Since we have removed the exponents causing decoherence, we naively expect the result to be a pure state, a vector that is a superposition of the four path branches. But, in such a vector there can only be three observable phases, while, in \eqref{eq:rhofnogamma} we have four such phases (global phases cancel in density matrices). The issue is that \eqref{eq:rhofnogamma} is \textit{not} a quantum mechanical state: it is easy to verify numerically for random values of the phases that it is not a positive matrix; it yields negative probabilities. To give an example, for the following set of values $(\varphi^{\tc{b}}_{\tc{al}},\varphi^{\tc{b}}_{\tc{ar}}, \varphi^{\tc{a}}_{\tc{bl}}, \varphi^{\tc{a}}_{\tc{br}})=(\frac{\pi}{3},\frac{\pi}{2},0,0)$, \eqref{eq:rhofnogamma} has one negative eigenvalue $-0.0653$. In the next section, we will see that this non-physicality is manifested as a violation of complementarity.

{\bf (ii).~Retro-causation in stationary phase approximation.} 
A different way of neglecting quantum fluctuations is to take a stationary phase approximation, which yields a proper density matrix. But, there is a subtler issue. This approximation of a quantum field theory  introduces retro-causation~\cite{eirini}. To see how this pathology is manifested in our context, we calculate again the final state of the masses
using the closed time path (CTP) formalism, a path-integral technique in which the field degrees of freedom are fiducially doubled~\cite{2021bentov}.  The CTP formalism allows to calculate expectation values of the form $\langle \omega^f_{\sigma'} |\omega^f_{\sigma}\rangle=\langle \omega| (\hat{U}^{\sigma'}_\phi (t_f,t_i))^{\dagger}\hat{U}^{\sigma}_\phi (t_f,t_i)|\omega\rangle$, whereas the usual path integral only calculates scattering amplitudes. 

The overlap of two evolved field states reads
\begin{equation}
\label{eq:toyOverlapsPI}
\langle \omega^f_{\sigma'} |\omega^f_{\sigma}\rangle  ={ \int }   \mathcal{D}\left[ \phi(\mf x) \right]\,  \mathcal{D}\left[ \phi'(\mf x) \right]\,  e^{ i  S_{CTP}\left[ \phi,\phi';J_{\sigma},J_{\sigma'} \right] } 
\end{equation}
with the action $S_{CTP}[\phi,\phi';J_{\sigma},J_{\sigma'}]$ given by
\begin{align}
\label{sbmveqSCTP}
\int  \dd^4\mf x\big( -\frac{1}{2}\partial_{\mu}\phi\partial^\mu \phi+\phi J_{\sigma} + \frac{1}{2}\partial_{\mu}\phi'\partial^\mu \phi'-\phi'J_{\sigma'} \big).
\end{align}
The two scalar field variables $\phi(\mf x)$ and $\phi'(\mf x)$ satisfy vacuum boundary conditions for $t=t_i$, and also coincide at $t=t_f$, $\phi(t_{f}, \boldsymbol{x})=\phi'(t_{f}, \boldsymbol{x})$. The external current for the interaction \eqref{eq:HAHB} is\footnote{For the toy model~\eqref{eq:artificialHI},
$J_{\sigma}(\mf x)=\sum_{a}(-1)^{s_a}\,\lambda_{a}(t)\,\delta^3(\boldsymbol{x}-\boldsymbol{x}_{a})$.}
\begin{equation}
\label{eq:Current}
J_{\sigma}(\mf x)=-\sum_{a}\lambda_{a}\,\delta^3(\boldsymbol{x}-\boldsymbol{x}^{a}_{\sigma}(t)).
\end{equation}
The path integral~\eqref{eq:toyOverlapsPI} is in Gaussian form. Defining the currents $J_q (\mf x)=J_{\sigma}(\mf x)-J_{\sigma'}(\mf x)$ and $J_{c}(\mf x)=\frac{J_{\sigma}(\mf x)+J_{\sigma'}(\mf x)}{2}$ the integration can be performed to get:
\begin{align}
\label{eq:Overlaps}
    \langle \omega^f_{\sigma'} |\omega^f_{\sigma}\rangle  & = \exp \Bigg(\iint \dd^4 \mf x \dd^4\mf y \Bigl[iJ_q (\mf x) \,G_{\tc r}(\mf x;\mf y) J_c(\mf y) \nonumber \\  & -\frac{1}{4} J_q (\mf x) \,H(\mf x;\mf y) J_q(\mf y) \Bigr]\Bigg).
\end{align}

The stationary phase approximation amounts to computing the path integral~\eqref{eq:toyOverlapsPI} on the classical field configurations $\phi_{\text{cl}}(\mf x)$, for which $\frac{\delta S_{CTP}}{\delta \phi}=\frac{\delta S_{CTP}}{\delta \phi'}=0$. Replacing the classical solutions $\phi_{\text{cl}}(\mf x)  = \int \dd^4 \mf y G_{\tc{r}}(\mf x ; \mf y) J_{\sigma}(\mf y)$ and $\phi'_{\text{cl}}(\mf x)  = \int \dd^4 \mf y G_{\tc{r}}(\mf x ; \mf y) J_{\sigma'}(\mf y)$ into~\eqref{sbmveqSCTP}, 
 the overlaps of the evolved field states in~\eqref{eq:toyOverlapsPI} are approximated as
\begin{equation}
\label{eq:toyOverlapsOS}
\begin{aligned}
\langle\omega_{\sigma'}^f| \omega_{\sigma}^f \rangle \approx \mathrm{exp} \Biggl\{\frac{i}{2} \iint  \dd^4\mf x \dd^4\mf y \Bigl( J_{\sigma}(\mf x)G_{\tc r}(\mf x;\mf y) J_{\sigma}(\mf y) \\
- J_{\sigma'}(\mf x)G_{\tc r}(\mf x;\mf y) J_{\sigma'}(\mf y)\Bigr) \Biggr\},
\end{aligned}
\end{equation}
see Appendix \ref{SPA} for details. This is a pure phase: there is no exponential suppression as $H$ does not appear. Explicitly, the final state \eqref{eq:rhofpath} is now approximated by
\begin{equation}
\hat{\rho}^f_{\tc{ab}}=\frac{1}{4}\left( \begin{matrix}
1  & e^{\frac{i}{2}(\varphi^{\tc b}_{\tc{al}} +\Delta_{\tc b}^{\tc{a}\tc{l}} )}   & e^{\frac{i}{2}(\varphi^{\tc a}_{\tc{bl}} +\Delta_{\tc a}^{\tc{b}\tc{l}}) } &  e^{\frac{i}{2} \Phi_1 } \\
& 1 & e^{\frac{i}{2} \Phi_2 } &  e^{\frac{i}{2}( \varphi^{\tc a}_{\tc{br}} + \Delta_{\tc a}^{\tc{b}\tc{r})} } \\
 &  & 1 & e^{\frac{i}{2}( \varphi^{\tc b}_{\tc{ar}} + \Delta_{\tc b}^{\tc{a}\tc{r}} ) } \\
  &  & & 1
\end{matrix} \right) 
\label{eq:rhofSP}
\end{equation}
where we defined
\begin{align}
    \Delta^{\tc{b}{s_\tc{b}}}_{\tc{a}} = \varphi^{\tc{b}{s_\tc{b}}}_{\tc{al}}- \varphi^{\tc{b}{s_\tc{b}}}_{\tc{ar}}\\
   \Delta^{\tc{a}{s_\tc{a}}}_{\tc{b}} = \varphi^{\tc{a}{s_\tc{a}}}_{\tc{bl}}- \varphi^{\tc{a}{s_\tc{a}}}_{\tc{br}}\\
    \Phi_1 = \varphi^{\tc b}_{\tc{ar}}+ \varphi^{\tc a}_{\tc{br}} + \Delta^{\tc{bl}}_{\tc{a}} + \Delta^{\tc{al}}_{\tc{b}}\\
    \Phi_2 = -\varphi^{\tc b}_{\tc{ar}}+\varphi^{\tc a}_{\tc{br}} + \Delta_{\tc a}^{\tc{b}\tc{r}} - \Delta_{\tc b}^{\tc{a}\tc{r}}
    \label{eq:phasesDefSP}
\end{align}

Contrary to the state \eqref{eq:rhofnogamma}, found in the first half of this section by imposing that $\Gamma_{\tc{a}}=\Gamma_{\tc{b}}=\Gamma_{\tc{ab}}=0$, \eqref{eq:rhofSP} is a proper quantum mechanical state. Despite appearances, only three linearly independent phases enter \eqref{eq:rhofSP}, and it is a pure state. Up to a global phase,\footnote{Due to diverging self-interactions this global phase is formally infinite. It cancels out at the level of the density matrices.} the corresponding state vector is
\begin{equation}
\label{eq:SPApurestate}
\ket{\Psi^f_{\tc{ab}}}=\frac{1}{2}\sum_{s_\tc{a},s_\tc{b}}  e^{\frac{i}{2}(\varphi^{\tc{a}s_{\tc{a}}}_{\tc{b}s_{\tc{b}}} + \varphi^{\tc{b}s_{\tc{b}}}_{\tc{a}s_{\tc{a}}})}\ket{s_{\tc{a}} s_{\tc{b}}}.
\end{equation}
 The stationary phase approximation allows to neglect quantum fluctuations without `breaking' quantum mechanics. However, by inspection of \eqref{eq:rhofSP}, something is wrong: we saw in Section \ref{sec:pathModel} that the phase combinations $\Delta_b^{a s_a}$ are not observable, the observable phases differences are the
 $\varphi^{\tc{a}}_{\tc{b}s_{\textsc{b}}}$
and
$\varphi^{\tc{b}}_{\tc{a}s_{\textsc{a}}}$
 defined in \eqref{eq:obsPhases}. 
 
To see the issue more clearly, we  examine the reduced density matrix of A
\begin{equation}
\label{eq:SPAreducedA}
\hat{\rho}_\tc{a}(t_f)= 
\frac{1}{2}\begin{pmatrix}
\;1\;  &  \frac{1}{2}(e^{ \frac{i}{2} (\varphi^{\tc a}_{\tc{bl}} +\Delta_{\tc a}^{\tc{b}\tc{l}})} + e^{  \frac{i}{2} (\varphi^{\tc a}_{\tc{br}} + \Delta_{\tc a}^{\tc{b}\tc{r}}) } ) \\ 
 & 1
\end{pmatrix} 
\end{equation}
It includes the phases  $\Delta_{\tc a}^{\tc{b}\tc{l}}$ and $\Delta_{\tc a}^{\tc{b}\tc{r}}$: these are \textit{sourced} by A, they depend on B's trajectories at \textit{later} times. The origin of this retro-causal behaviour is that in \eqref{eq:Overlaps}, when $J_q(\mf x)G_\tc{r}(\mf x; \mf y) J_c(\mf y)$ is expanded it include the term $J_\sigma (\mf x) E(\mf x;\mf y) J_{\sigma'}(\mf y)$, where $E$ is the so-called causal propagator, defined by $E(\mf x;\mf y)=i\langle [\hat{\phi}(\mf x),\hat{\phi}(\mf y)]\rangle $). In detail,
\begin{equation}
\begin{aligned}
    2 \, J_q(\mf x)G_\tc{r}(\mf x; \mf y) J_c(\mf y) & =  J_{\sigma}(\mf x)G_{\tc r}(\mf x;\mf y) J_{\sigma}(\mf y) \nonumber \\ &- J_{\sigma'}(\mf x)G_{\tc r}(\mf x;\mf y) J_{\sigma'}(\mf y) \nonumber \\ & +  J_\sigma (\mf x) E(\mf x;\mf y) J_{\sigma'}(\mf y). 
\end{aligned}
\end{equation}
When we take the stationary approximation, we effectively keep only the first two lines of the above expression. That is, in this case (ii) we force  $E(\mf x;\mf y)=H(\mf x;\mf y)=0$, while in case (i) we were only setting $H(\mf x;\mf y)=0$. The causal propagator allows for quantum superpositions of past and future propagation, which is a quantum-field-theoretic feature not present in the classical theory and is hence neglected in the stationary phase approximation (recall that $E= G_{\tc{r}} - G_{\tc{a}}$). As we see in a moment, the approximate state \eqref{eq:rhofSP} describes spacetime local generation of entanglement,
allowing for signalling at spacelike separation~\cite{eirini}.

\section{The paradox}

\label{sec:Paradox}
We now turn to a thought-experiment presented in~\cite{2018belenchiaQuantumSuperpositionMassive, 2019belenchiaInformationContentGravitational, 2022danielsonGravitationallyMediatedEntanglement}, in which the pathologies discussed above can be given a clearer physical interpretation. It goes briefly as follows. Two particles A and B can be manipulated by agents, which we also call A and B. A acts as a source and B as a test particle living in the gravitational field of A. Mass A is path-split in the far past, and B lives in its gravitational field. Until some time $t_r$, the interaction of particle B with the field is always `turned off' (the initial time $t_i$ is formally in the far past). This can be imagined to happen, for instance, because of particle B being placed in a strong trap (a `box'), that overwhelms the gravitational interaction.
 
Between times $t_r$ and $t_f$, agent A begins recombining her particle and subsequently does interferometric measurements (e.g.~measuring on the complementary basis of the paths), with the aim to check the coherence of her particle (the protocol is repeated many times, in order to gather statistics). Between times $t_r$ and $t_f$, B has two options: to release the particle from the trap, or not. If he does, the particle starts feeling the gravitational field of A. During times $t_r$ and $t_f$, A and B are in spacelike separation. If one imagines that particle B lives in a superposition of fields sourced by A, then B will get entangled with the field, by falling in a superposition of paths. Therefore, the argument goes, since B can acquire which-path information by measuring the path, and B is being traced out when A locally checks for coherence, A should decohere. This is a paradoxical situation, since it clearly allows for instantaneous signalling between A and B, assuming both have been made aware of the aforementioned protocol: B can choose to open the box or not, therefore, A can know what B chose to do by checking if her particle has decohered (possibly partly, and assuming no other sources of decoherence).

\begin{figure}
    \centering
\includegraphics[scale=0.75]{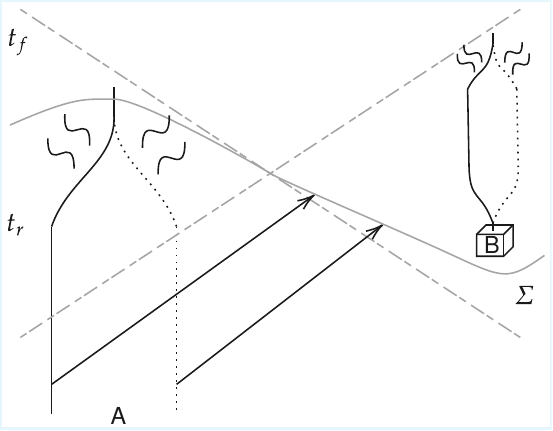}
    \caption{The set--up of the paradox \cite{2016mariExperimentsTestingMacroscopic,2018belenchiaQuantumSuperpositionMassive,2019belenchiaInformationContentGravitational,2022danielsonGravitationallyMediatedEntanglement}. Mass A is in a path--superposition in the distant past $t_i$. Then, it undergoes a recombination process at $t_r$. The recombination of mass A is spatially separated from the splitting and recombination of mass B. The dashed lines indicate null rays. $\Sigma$ is a hypersurface for which A has recombined but B has not started the protocol. Which-path information propagates causally according to the arrows, and is encoded in the pure phases $\varphi^{\tc{b}}_{\tc{a}s_\tc{a}}$. Radiation that causes decoherence also propagates causally (wriggly lines) and is produced locally to A and B. The amount of this radiation is encoded in the exponentially suppressing exponents $\Gamma_\tc{a}$ and $\Gamma_\tc{b}$. }
    \label{fig:paradox}
\end{figure}

We now discuss a subtlety: our analysis \textit{considers a variation of the above paradox}. In the models we have studied, B is not a test particle falling in a gravitational field in superposition sourced by A. Instead, we are taking (1) both particle A and B to be massive (2) both set in superposition by some external means, see Figure \ref{fig:paradox}. Therefore, the phases being created between them are of Aharonov-Bohm type, in the approximate sense that the force is negligible (the particles are not `falling' towards each other, the phases are solely due to potentials). This is the main insight exploited in table-top entanglement-through-gravity protocols. In order to judge the relevance of the paradox for said proposals---the task we have set ourselves here---it is necessary to study it within this regime. The analogue of B `opening the box' is to allow B to choose whether to make his particle couple to the field or not. 
This is similar to the toy model of Section \ref{sec:Scalar}, where had an artificial time dependence in the coupling constant, see \eqref{eq:artificialHI}. Analogously, for the model with the path superposition of Section \ref{sec:pathModel}, we can consider a time dependent coupling constant for B in \eqref{eq:HAHB}, taking it zero (`switched of') until time $t_r$. At that time, B can choose to couple to the field with a constant coupling $\lambda_\tc{b}$ (to `open the box') or keep the coupling to zero. 

 Since A is in spacelike separation with B when the latter will begin the protocol, no quantum phases are sourced by B. That is, 
\begin{equation}
\varphi^{\tc{a}}_{\tc{b}s_\tc{b}}= 0.
\end{equation}
Next, we introduce the visibility $\mathcal{V}_\tc{a}$ of A and the distingushability $\mathcal{D}_{\tc b}$ of (A's state as seen by) B. Quantum theory demands that \cite{2022hidaka,2023hidaka, Uncertainty1,Uncertainty2}:
\begin{equation}
\label{eq:complviolation}
    \mathcal{V}^2_\tc{a}+ \mathcal{D}^2_\tc{b} \leq 1
\end{equation}
This is the definition of \textit{complementarity} in this context. The visibility of A is given by $\mathcal{V}_\tc{a}=2|\bra{L} \hat{\rho}_\tc{a}(t_f)\ket{R}_\tc{a}|$, it encodes the coherence of its reduced  state. The distinguishability of B is given by $
\mathcal{D}_{\tc b}=\frac{1}{2} \mid\mid \hat{\rho}^{\tc l}_{\tc b}-\hat{\rho}^{\tc r}_{\tc b} \mid\mid_{1}$. It encodes `how much' B can tell apart the two paths of A, where $\mid\mid \cdot \mid\mid_{1}$ is the trace norm. The matrix elements $\hat{\rho}^{\tc{l}/\tc{r}}_{\tc b}$ are the `state' of B when A follows the left/right path and can be directly read from the final state $\hat{\rho}_\tc{ab}(t_f)$, they are its upper left/lower right components. They are 
are given by
\begin{equation}
\begin{aligned}
\hat{\rho}^{\tc l}_{\tc b}&=2\,\langle L |\hat{\rho}_{\tc{ab}}(t_f)  | L \rangle_{\tc a} \\
\hat{\rho}^{\tc r}_{\tc b}&=2\,\langle R |\hat{\rho}_{\tc{ab}} (t_f) | R \rangle_{\tc a}. 
\end{aligned}
\end{equation}
We have that $0\leq \mathcal{V}_{\tc a}\leq 1$ and $0\leq \mathcal{D}_{\tc b}\leq 1$. The expressions for $\mathcal{V}_{\tc a}$ and $\mathcal{D}_{\tc b}$ given in the rest of this Section are derived in Appendix~\ref{sec:AppVisDis}. We are now ready to see how the paradox arises.

\smallskip

{\bf (i). Complementarity violation when neglecting quantum fluctuations by hand.} If we neglect quantum fluctuations by setting to zero the Hadamard function, which amounts to imposing
\begin{equation}
\label{eq:qFluctsZero}
  \Gamma_\tc{a}=\Gamma_\tc{b}=0,
\end{equation}
the reduced state of A becomes the constant matrix
\begin{equation}
\label{eq:Areducednogamma}
    \hat{\rho}_\tc{a}(t_f)= 
    \frac{1}{2}\begin{pmatrix}
    1  &  1\\ 
    1 & 1
    \end{pmatrix}.
\end{equation}
Clearly, no-signalling is not violated: what takes place locally to A is independent of any choices of B. 

However, complementarity is violated. The distinguishability of B is
\begin{equation}
\label{eq:DBGamma=0}
    \mathcal{D}_{\tc b}=  \left| \sin\left(\frac{\varphi^{\tc b}_{\tc{al}}-\varphi^{\tc b}_{\tc{ar}}}{2} \right)\right| \geq0
\end{equation}
and from~\eqref{eq:Areducednogamma} we have $\mathcal{V}_{\tc a}=1$. Thus, we get $\mathcal{V}^2_\tc{a}+ \mathcal{D}^2_\tc{b}\geq 1$ whenever $\mathcal{D}_{\tc b}\neq 0$, which will necessarily happen for some values of the phases. Intuitively, this violation of complementarity \eqref{eq:complviolation} means that while B acquires which-path information about A---which should have `collapsed' her---she has remained coherent.

\smallskip

\smallskip

{\bf (ii). Violation of no-signaling in stationary phase approximation.} 
If the final state of A depends on the choice of B to couple to the field ($\lambda_\tc{b} \neq 0$) or not  ($\lambda_\tc{b}=0$) at time $t_r$, A and B can signal at spacelike separation. We may write this as occurring when it is true that
\begin{equation}
\label{eq:superluminal}
    \hat{\rho}_\tc{a}(t_f) [\lambda_\tc{b}=0] \neq \hat{\rho}_\tc{a}(t_f) [\lambda_\tc{b}\neq 0].
\end{equation}
 In the stationary phase approximation, the reduced state of A for the set-up of Figure~\ref{fig:paradox} is
\begin{equation}
\label{eq:ParadoxSPAreducedA}
\hat{\rho}_\tc{a}(t_f)= 
\frac{1}{2}\begin{pmatrix}
\;\;1  &  \frac{1}{2}(e^{ \frac{i}{2} \Delta^{\tc{bl}}_{\tc{a}} } + e^{  \frac{i}{2} \Delta^{\tc{br}}_{\tc{a}} } ) \\ 
 & 1
\end{pmatrix} .
\end{equation}
where recall that $\Delta^{\tc{b}{s_\tc{b}}}_{\tc{a}} = \varphi^{\tc{b}{s_\tc{b}}}_{\tc{al}}- \varphi^{\tc{b}{s_\tc{b}}}_{\tc{ar}}$, defined in \eqref{eq:phasesDefSP}, are unobservable phases that appear when taking the stationary phase approximation, see \eqref{eq:SPAreducedA} and the discussion following it. Had agent B chosen not to release his particle, i.e.~$\lambda_\tc{b}=0$, the state of A would instead be
\begin{equation}
\hat{\rho}_\tc{a}(t_f)= 
\frac{1}{2}\begin{pmatrix}
1  &  1\\ 
1 & 1
\end{pmatrix}.
\end{equation}
This allows for superluminal signalling, since \eqref{eq:superluminal} holds.

On the other hand, complementarity is satisfied. The visibility of A and distinguishability of B are given by
    \begin{equation}
    \label{eq:SPAVD}
        \begin{aligned}
            \mathcal{V}_\tc{a}&=  \left| \cos\left(\frac{\varphi^{\tc b}_{\tc{al}}-\varphi^{\tc b}_{\tc {ar}}}{4} \right)\right|\\
            \mathcal{D}_{\tc b}&=  \left| \sin\left(\frac{\varphi^{\tc b}_{\tc{al}}-\varphi^{\tc b}_{\tc {ar}}}{4} \right)\right|
        \end{aligned}
    \end{equation}
which implies $\mathcal{V}_\tc{a}^2 +  \mathcal{D}_{\tc b}^2 =1$.

\section{Entanglement through Newtonian interaction does not imply gravitons}
\label{sec:NoPertinenceNewton}
The violation of no-signalling analysed in the previous two Sections arises in a relativistic setting, where phases are created with retardation: it should not be confused with the trivial violation of no-signaling which arises when assuming a Newtonian action-at-a-distance.

In the Newtonian regime\footnote{The final state in the Newtonian limit is given in Appendix~\ref{sec:NewtonianLimit}.} there are no light-cones, and therefore no distinction between phases sourced by A or phases sourced by B. This approximation corresponds to demanding
\begin{equation}
\label{eq:NewtonianCondition}
\varphi^{\tc{a}s_\tc{a}}_{\tc{b}s_\tc{b}} \sim \varphi^{\tc{b}s_\tc{b}}_{\tc{a}s_\tc{a}} \equiv \varphi_{s_\tc{a}s_\tc{b}}
\end{equation} 
where we have introduced the notation $\varphi_{s_\tc{a}s_\tc{b}}=\varphi_{s_\tc{b}s_\tc{a}}$ temporarily, to emphasize that there is no longer a distinction between lower (source) and upper (recipient) indices. If B has coupled to the field,  the reduced density matrix of A is
\begin{equation*}
\label{eq:noGammareducedA}
2 \hat{\rho}_\tc{a}(t_f)= 
\begin{pmatrix}
\;\;1  &   \ \  \frac{1}{2}(e^{ i (\varphi_{\tc{ll}} -\varphi_{\tc{rl}} )} + e^{ i (\varphi_{\tc{lr}} -\varphi_{\tc{rr}}) } ) \\ 
 & 1.
\end{pmatrix} 
\end{equation*}
While, if B did not couple to the field, A's state is
\begin{equation*}
2 \hat{\rho}_\tc{a}(t_f)= 
\begin{pmatrix}
1  &  1 \\ 
1 & 1
\end{pmatrix} 
\end{equation*}
This allows for superluminal signalling, as \eqref{eq:superluminal} holds. Agent A can check with local to her interferometric experiments which of the above states she has (the latter is fully coherent and the former is not). So, B can communicate to A whether he has chosen to open the box or not.

 It is tempting to think of this as the `other side' of the paradox, in place of the case (ii). Recall that case (i) is when complementarity fails while no-signalling holds. In a sense, it seems that by introducing direct action-at-a-distance `complementarity has been salvaged'. Indeed, it is easy to check that $\mathcal{V}^2_{\tc a}+\mathcal{D}_{\tc b}^2=1$  when \eqref{eq:NewtonianCondition} holds. But this is unsurprising:  this regime corresponds to standard non-relativistic quantum mechanics for two particles in a potential; there is no light-cone structure.

A more abstract way to see that quantum fluctuations are not relevant if entanglement is assumed to be created by a Newtonian interaction, is through the Robertson-Schr\"odinger inequality \cite{Uncertainty1}. For the model studied in Section \ref{sec:pathModel}, it reads
\begin{equation}
\label{eq:RSineq}
\Gamma_{\tc{a}}\Gamma_{\tc{b}}\geq\left((\varphi^{\tc b}_{\tc{ar}}-\varphi^{\tc a}_{\tc{br}}) - (\varphi^{\tc b}_{\tc{al}}-\varphi_{\tc{bl}}^{\tc a})\right)^2
\end{equation}   
see Appendix~\ref{sec:AppUnc} for the derivation. This is an uncertainty relation between the exponents encoding quantum fluctuations and the phases encoding causal propagation of which-path information. For the setup of the thought experiment, since $\varphi^{\tc{a}}_{\tc{b}s_\tc{b}}= 0$, it comes down to
\begin{equation}
\label{eq:uncertainty}
\Gamma_{\tc{a}}\Gamma_{\tc{b}} \geq \left(\varphi^{\tc b}_{\tc{ar}} - \varphi^{\tc b}_{\tc{al}}\right)^2
\end{equation}   
 When this inequality holds, it guarantees complementarity, in the sense of \eqref{eq:complviolation}. When it is violated, complementarity may or not hold.  Independently on whether we neglect quantum fluctuations ad-hoc or through a stationary approximation, the inequality becomes
\begin{equation}
\label{eq:uncertaintyGamma=0}
0 \geq \left(\varphi^{\tc b}_{\tc{ar}} - \varphi^{\tc b}_{\tc{al}}\right)^2 
\end{equation}
It is agnostic on what has been broken, it is violated both for case (i) and for case (ii), by the same amount.\footnote{Note also that $\Delta^{\tc{br}}_{\tc{a}}-\Delta^{\tc{bl}}_{\tc{a}} = \varphi^{\tc b}_{\tc{ar}} - \varphi^{\tc b}_{\tc{al}}$. } 

On the other hand, it is easy to be convinced that demanding that complementarity holds, while neglecting  quantum fluctuations, implies that signalling has been introduced. Neglecting quantum fluctuations while complementarity holds, implies that A is not fully coherent (since the state of B has which path information). But, since there are no quantum fluctuations, the only way that A could have decohered is through a dependence to the coupling of B, which the agents can exploit to signal at spacelike separation. The paradoxical situation of the thought experiment exemplifies that either complementarity or no-signalling must break when the Robertson inequality is \textit{violated}. 

On the contrary, if we assume an instantaneous generation of entanglement the Robertson inequality \textit{always holds}. The right hand side of \eqref{eq:RSineq} collapses to zero  due to \eqref{eq:NewtonianCondition}: any value of quantum fluctuations is compatible with it, including zero. This simply confirms that there is nothing contradictory about neglecting quantum fluctuations of the field in the Newtonian limit.

\section{No spacelike influences, No duality between gravitons and Newtonian field}
\label{sec:NoSpacelikeInfluences}
A question that seems to have caused confusion in the literature, is whether `spacelike influences' between A and B play a role in the interpretation of the paradox, a point discussed e.g.~in \cite{2022danielsonGravitationallyMediatedEntanglement}. Spacelike correlations, generically present in the state of the field, are encoded by the non-local Hadamard terms $\Gamma_{\tc{ab}}$, which do not enter the thought experiment, see Section~\ref{sec:propagators} and the discussion above~\eqref{eq:reducedA}. There
are two distinct mechanisms of entanglement generation. The first is due to the causal interaction of the masses through the retarded propagators. This is the mechanism of entanglement generation relevant for `table top' protocols for quantum gravity---including in the relativistic regime where entanglement is generated with retardation. This type of entanglement  does not depend on the quantum fluctuations of the field. 

The second mechanism for entanglement generation is entanglement harvesting, a phenomenon fully due to correlations at spacelike separation that correspond to pre-existing entanglement in the field vacuum. These quantum fluctuations are encoded in the exponents $\Gamma_{\tc{ab}}$, which do not enter the reduced density matrices of A and B. Entanglement harvesting is a fundamentally different mechanism for entanglement generation than the entanglement discussed in the context of table-top quantum gravity. The quantum fluctuations that do enter the discussion of the paradox are the $\Gamma_\tc{a}$ and $\Gamma_\tc{b}$: these cause local decoherence and do not create any sort of entanglement.

A related point raised in \cite{2022danielsonGravitationallyMediatedEntanglement} is that the thought-experiment implies a `duality' between the radiation emitted due to decoherence and the entangling phases, depending on what foliation is chosen. We have not seen any evidence for this, in the models that we have studied here and solved for the final state of the probes, the two effects are distinct. 

This can be seen also more generally: with minimal assumptions for the probe-field interactions (a) the reduced state of A cannot be affected by whether B decides to couple to the field or not (b) the reduced state of B cannot be affected by A's recombination process. 

Let us see this in more detail (a similar argument has been presented in \cite{sudarsky}).  Let $U_{\textsc{a}+\textsc{b}}$ denote the joint interaction of A and B with the field. If B does not intersect the causal past of A's interaction with the field (while B is coupled to the field), we expect that the joint interaction factorises as $U_{\textsc{a}+\textsc{b}}= U_{\textsc{b}} U_{\textsc{a}}$.\footnote{Assuming that the local $S$-matrices are well defined and that they satisfy causal factorisation. For a proof of causal factorisation for Unruh-DeWitt-type detector-field interactions see \cite{PhysRevD.103.085002}.} This has the consequence that 
\begin{equation}
    \rho_{\textsc{a}}= \text{tr}_{\textsc{b}\textsc{f}} \left( U_{\textsc{a}+\textsc{b}} \rho_0  U^{\dagger}_{\textsc{a}+\textsc{b}} \right) = \text{tr}_{\textsc{f}} \left(  U_{\textsc{a}} \rho_{\textsc{af}} U^{\dagger}_{\textsc{a}}  \right)
\end{equation}
where $\rho_0= \rho_{\textsc{af}}\rho_{\textsc{b}}$ is the total initial state of field (F), A and B (assuming that B is initially uncorrelated with A and the field). This implies (a), the amount of coherence encoded in A's state after she begins the recombination process at $t_r$, does not depend on whether B has interacted with the field or not.

Regarding point (b), we can assume that A's total interaction with the field can be decomposed with respect to a spatial hypersurface that passes through her interaction region and separates the recombination process (rec) from what comes before (split), namely $U_{\tc{a}}= U_{\text{rec}}U_{\text{split}}$.\footnote{A proof that this holds in a similar setup to ours (quantum-controlled linear interactions between a scalar field and probes that have no internal dynamics) is given in \cite{simmons2026factorisation}.} Given that B is in a path superposition while spacelike separated from A's recombination process, this implies that 
\begin{equation}
    \rho_{\textsc{b}}= \text{tr}_{\textsc{a}\textsc{f}} \left(  U_{\textsc{b}}  U_{\text{split}} \rho_0  U^{\dagger}_{\text{split}}   U^{\dagger}_{\textsc{b}}\right).
\end{equation}
Then, B's correlations with A through the field can only depend on the part of A's splitting process (that intersects B's causal past) and not by radiation that she will emit during the recombination process. A more detailed derivation of this argument can be found in Appendix \ref{appendix: reduced states}. If spacelike influences were to play any role in the experiment, at least one of the above assumptions, must be lifted: for instance, by introducing an action-a-distance. 

We can now summarize how the conclusions arising from our analysis differ from that of~\cite{2022danielsonGravitationallyMediatedEntanglement}. If a spacelike hypersurface $\Sigma$ is drawn such that it is in the past of B opening the box, and in the future of A after having recombined, at first it might appear that there is no explanation why later on B will receive which-path information, see Figure \ref{fig:paradox}. Then, taking the point of view of the evolution with respect to a foliation that includes $\Sigma$, it may seem that B must somehow be interacting at spacelike separation with entangling radiation emitted by A. However, the premise is false. B \textit{will} receive the which path information \textit{causally} in such a foliation: the field on the hypersurface $\Sigma$ is in a quantum superposition, and encodes two phases $\varphi^{\tc b}_{\tc{ar}}$ and $\varphi^{\tc b}_{\tc{al}}$. These are `stuck' on that hypersurface at the moment in time defined by it. However that foliation is continued in the future, these phases will reach B, thereby B will acquire which-path information by the same mechanism as in any other foliation. 

\section{Conclusions}\label{sec:conc}
We studied a paradox in quantum field theory coupled to two distant quantum masses that arises when quantum fluctuations are neglected and has drawn interest for its potential to support a major epistemological significance for entanglement-through-gravity experiments. We critically assessed the degree to which this is the case. Our main result is to show that the tension between complementarity and no-signalling concerns two distinct ways of taking the approximation of negligible quantum fluctuations: (i) directly setting the exponentially suppressing exponents in the final state to zero, which breaks complementarity, and (ii) taking a stationary phase approximation, which breaks no-signalling. In both cases, entanglement arises locally in spacetime, due to phases that are created with retardation. We discussed that case (ii) should not be conflated with the trivial violation of causality that arises in the Newtonian limit, where no light-cone structure is present: the paradox is firmly placed in a setting where spacetime locality holds.

We commented on the idea that the effect might be viewed from different perspectives, via different foliations of the relevant spacetime region, so that a party acquires information on the state of another either from the `Newtonian part' or the `radiative part' of the field \cite{2022danielsonGravitationallyMediatedEntanglement,2018belenchiaQuantumSuperpositionMassive}. In our analysis, there is no `duality' between the quantum fluctuations that cause the decoherence and the retarded propagators that cause the entanglement between A and B. The quantum fluctuations that give rise to this radiation are produced locally to A or to B and introduce local decoherence: they do \textit{not} generate entanglement. This decohering radiation produced by the particles would propagate \textit{causally}, and would only reach the parties well after the completion of the protocol, see Figure \ref{fig:paradox}.

Relatedly, it appears to be a received view that decoherence between the parties takes place non--causally, via spacelike fluctuations. We clarified that nothing of the sort takes place in the thought-experiment. Spacelike quantum fluctuations, while present in the full state, do not appear in the reduced density matrices of the parties. They also do not relate to the kind of entanglement that would be seen in table-top experiments---even if retarded generation of entanglement was observed. Rather, spacelike quantum fluctuations are the kind of correlations that are studied in the context of entanglement harvesting, a phenomenon which plays no role in the thought experiment. 

The epistemological relevance of entanglement-through-gravity in the Newtonian regime has been much discussed. One common line of argument is to assume that the interaction
is in some sense ‘mediated’ by gravity, and conclude that the gravitational
field must be able to carry quantum information,  \cite{2017boseSpinEntanglementWitnessa, 2017marlettoGravitationallyinducedEntanglementTwo, 2022galley, 2023galley}. This has been criticised, as it is far from clear in what sense there are `mediating degrees of freedom'---even when assuming spacetime local interactions \cite{Christodoulou:2022knr,DiBiagio:2023jiu,Fragkos:2022tbm,eirini}. As a result, a lively debate on the importance of detecting entanglement-through-gravity in the Newtonian regime is taking place, e.g.~\cite{2023huggett,Aziz:2025ypo,Trillo:2025kio,Bose:2022uxe}. 

An intuitive interpretation of the effect in the Newtonian regime is as spacetime being set in superposition \cite{Christodoulou:2018cmk,Bengyat:2023hxs}. In the language of this work, this can be seen from the fact that the entangling phases are the Newtonian limit of retarded Green's functions. The latter appear in the final state as pure phases, they correspond to a quantum superposition of the macroscopically distinct field configurations sourced by the alternative paths of the masses. Therefore, the entangling phases in entanglement-through-gravity protocols---including when entanglement is created with retardation---depend only on the classical equations of motion: they are not quantum fluctuations. In other words, \textit{coherent information can be exchanged in a relativistic setting without exchanging quanta}.

The thought experiment adds to the epistemological importance of entanglement-through gravity, but only in a context where entanglement is created with retardation. Therefore, in order to employ the paradox as a consistency argument, it seems necessary that retardation effects in entanglement-through-gravity are detected, not assumed, for the following reasons: (a) There is nothing contradictory if Newtonian entanglement is detected and gravitons do not exist.
(b) To assume spacetime-local generation of entanglement is a different causality assumption than to request that no-signalling holds: one side of the paradox is precisely that while entanglement is created with retardation, no-signalling is violated. (c) While experiments that could detect relativistic corrections due to retardation in entanglement-through-gravity \cite{Christodoulou:2022mkf,Bengyat:2023hxs}
 are significantly more difficult than detecting Newtonian entanglement, they should be far easier than a direct detection of gravitons (which may be impossible \cite{Dyson:2013hbl,Rothman:2006fp}). 

In summary, the crucial input for the paradox is \textit{retarded generation of entanglement}, according to the light-cone structure. This notion of causality is distinct from the principle of no-signalling and should be demonstrated by experiment. Detection of spacetime local generation of entanglement-through-gravity between two spacelike separated parties, in conjuction with the demand that complementarity and no-signaling holds, provides evidence for the existence of gravitons.

\begin{acknowledgements}
We acknowledge support from the ID\# 62312 grant from the John Templeton Foundation, as part of the `The Quantum Information Structure of Spacetime’ Project (QISS). The opinions expressed in this project are those of the authors and do not necessarily reflect the views of the John Templeton Foundation. MP acknowledges that this research was funded in part by the Austrian Science Fund (FWF)
[10.55776/F71] and [10.55776/COE1]. TRP is thankful for financial support from the Olle Engkvist Foundation (no.225-0062) and to the TURIS-IQOQI Long-Term Fellowship, which allowed him to become a part of this collaboration during his time in Vienna. This work was partially conducted while TRP was still a PhD student at the Department of Applied Mathematics at the University of Waterloo, the Institute for Quantum Computing, and the Perimeter Institute for Theoretical Physics. TRP acknowledges partial support from the Natural Sciences and Engineering Research Council of Canada (NSERC) through the Vanier Canada Graduate Scholarship. Nordita is partially supported by Nordforsk.
\end{acknowledgements}

\bibliography{references}

\onecolumngrid

\appendix

\section{Non-perturbative calculation of the joint state after the interaction with the field} \label{appendix: Magnus}

In  this appendix we solve the dynamics of the field-masses system non-perturbatively, using the Magnus expansion. We derive the reduced density matrix of the two masses for the toy-model of Section~\ref{sec:Scalar} and the path-superposition model of Section~\ref{sec:pathModel}, corresponding to equations \eqref{eq:toyrhofFull} and \eqref{eq:rhofpath} of the main text respectively. The main steps proceed as follows. The Magnus expansion yields phases that correspond to self-interacting terms, these drop out of the final state of the masses. Then, using the Weyl relations we get combinations of phases that depend on the field's causal propagator. When evaluating over the state $\omega$ we get phases that depend on the field's two-point function, whose imaginary part (the Hadamard function $H$) encodes the quantum fluctuations and leads to decoherence.

\subsection*{Toy model}

We start with the interaction Hamiltonian density of the toy-model~\eqref{eq:artificialHI}, which can be written as
\begin{equation}
     \hat{{h}}_{int}(\mf x) = -\Lambda_{\tc{a}}(\mf x)\hat{\sigma}_{\tc{a}}\hat{\phi}(\mf x)-\Lambda_{\tc{b}}(\mf x)\hat{\sigma}_{\tc{b}}\hat{\phi}(\mf x).
\end{equation}
where $\hat{\sigma}_a=\ket{0}\!\!\bra{0}_a-\ket{1}\!\!\bra{1}_a$ is the third Pauli matrix acting on the spin $a=\text{A},\text{B}$ and $\Lambda_a(\mf x)=\lambda_a(t)\delta^3(\bm x-\bm x^a)$ is a smearing function that includes the switching $\lambda_a(t)$ and enforces the interaction to be local at the position of each mass. In what follows we define the smeared field operator as $\hat{\phi}_a(\mf x)=\Lambda_a(\mf x) \hat{\phi}(\mf x)$.

In order to solve the dynamics of states evolving under the Hamiltonian above, we can use the Magnus expansion. Notice that due to the canonical commutation relations satisfied by the scalar field, we have $[[\hat{h}(\mf x),\hat{h}(\mf x')],\hat{h}(\mf x'')]=  0$. In this case, the Magnus expansion allows us to write the time evolution operator as
\begin{equation}
\label{eq:AppMagnus}
    \hat{U}_I = e^{\hat{\Theta}_1+\hat{\Theta}_2},
\end{equation}
where
\begin{align}
    \hat{\Theta}_1 &= - i \int \dd V \hat{h}_{int}(\mf x) =  i \int \dd V \left( \hat{\phi}_{\tc{a}} (\mathsf{x}) \hat{\sigma}_{\tc{a}} +  \hat{\phi}_{\tc{b}} (\mathsf{x}) \hat{\sigma}_{\tc{b}}\right)\label{eq:TH2}\\
    \hat{\Theta}_2 &= - \frac{1}{2} \int \dd V \dd V' \theta(t-t')[\hat{h}_{int}(\mf x), \hat{h}_{int}(\mf x')] \nonumber
    \\&=- \frac{1}{2} \int \dd V \dd V' \theta(t-t')[ \hat{\phi}_{\tc{a}} (\mathsf{x}) \hat{\sigma}_{\tc{a}}+  \hat{\phi}_{\tc{b}} (\mathsf{x}) \hat{\sigma}_{\tc{b}}, \hat{\phi}_{\tc{a}} (\mathsf{x}') \hat{\sigma}_{\tc{a}}+  \hat{\phi}_{\tc{b}} (\mathsf{x}') \hat{\sigma}_{\tc{b}}] \nonumber\\
    &= \frac{1}{2}\Big(i \mathcal{G}_{\tc{a}} +  i \mathcal{G}_{\tc{b}}+i \Delta_{\tc{a}\tc{b}}\hat{\sigma}_{\tc{a}}\hat{\sigma}_{\tc{b}}\Big).
\end{align}
The terms
\begin{equation}
    \mathcal{G}_a=  \int \dd V \dd V' \theta(t-t')[  \hat{\phi}_{{a}} (\mathsf{x}),  \hat{\phi}_{{a}} (\mathsf{x}')]
\end{equation}
are pure phases that, while formally infinite, will drop out from the expression of the reduced states. Performing the spatial integration we get $ \mathcal{G}_a=  \int \dd t \dd t' \lambda_a(t) G_{\tc{r}}(t, \bm{x}^a; t', \bm{x}^a) \lambda_a(t')$. The term $\Delta_{\tc{a}\tc{b}}$ is an integral of the symmetric propagator
\begin{equation}
    \Delta_{\tc{a}\tc{b}}=   \int \dd t \dd t' \lambda_\tc{a}(t) (G_{\tc{r}}(t, \bm{x}^\tc{a}; t', \bm{x}^\tc{b})+G_{\tc{a}}(t, \bm{x}^\tc{a}; t', \bm{x}^\tc{b})) \lambda_\tc{b}(t')=G^{\tc{a}\tc{b}}_{\tc r} + G^{\tc{a}\tc{b}}_{\tc{a}}.
\end{equation}

We can now rewrite our time evolution operator as a product of exponentials of different kinds of operators
\begin{equation}\label{eq:UI2g}
    \hat{U}_{int} = e^{ i  \hat{\phi}_{\tc{a}} \hat{\sigma}_{\tc{a}} +i \hat{\phi}_{\tc{b}} \hat{\sigma}_{\tc{b}}} e^{ i \mathcal{G}_{\tc{a}}+ i \mathcal{G}_{\tc{b}}} e^{ \frac{i}{2} \Delta_{\tc{a}\tc{b}}\hat{\sigma}_\tc{a}\hat{\sigma}_\tc{b}},
\end{equation}
where $ \hat{\phi}_a= i \int \dd V  \hat{\phi}_{a} (\mathsf{x})$ for $a=\text{A,B}$.
Notice that the first exponential above contains a direct interaction of each of the masses with the field, the second term only contains local interactions experienced by mass A and B individually corresponding to global phases that do not affect the overall dynamics, and the last term contains symmetric exchange of information between masses A and B through the scalar field. We decompose the unitary time evolution operator $\hat{U}_{int}$ as a product of commuting unitaries, $\hat{U}_{int} = \hat{U}_\tc{c}\hat{U}_\phi$, where
\begin{equation}
    \hat{U}_\phi = e^{ i  \hat{\phi}_{\tc{a}}  \hat{\sigma}_{\tc{a}} + i  \hat{\phi}_{\tc{b}}  \hat{\sigma}_{\tc{b}}} ,
\end{equation}
which contains all of the dependence in the scalar field, and
\begin{equation}
\label{U_c}
    \hat{U}_\tc{c}= e^{ i \mathcal{G}_{\tc{a}}+ i \mathcal{G}_{\tc{b}}} e^{ \frac{i}{2} \Delta_{\tc{a}\tc{b}}\hat{\sigma}_\tc{a}\hat{\sigma}_\tc{b}},
\end{equation}
which does not directly depend on the degrees of freedom of the scalar field. 

In order to resolve the dynamics of the masses, we must trace over the state of the field. To do so, we focus first on the unitary $\hat{U}_\phi$. Let us fix an initial state $\hat{\rho}_0 = \hat{\rho}_{\tc{a},0}\otimes \hat{\rho}_{\tc{b},0}  \otimes \hat{\rho}_\omega$. We then have
\begin{align}
    \hat{U}_\phi \hat{\rho}_0\hat{U}_\phi^\dagger &= e^{ i  \hat{\phi}_{\tc{a}} \hat{\sigma}_{\tc{a}} + i  \hat{\phi}_{\tc{b}} \hat{\sigma}_{\tc{b}}}\hat{\rho}_{0}e^{ -i  \hat{\phi}_{\tc{a}}  \hat{\sigma}_{\tc{a}} - i  \hat{\phi}_{\tc{b}}  \hat{\sigma}_{\tc{b}}}\nonumber\\
    &=\frac{1}{4}\sum_{{\substack{{}_{s_{\tc{a}}s_{\tc{b}} }\\{}_{s'_{\tc{a}} s'_{\tc{b}}}}}}  e^{ i s_{\tc{a}}  \hat{\phi}_{\tc{a}} + i s_{\tc{b}}  \hat{\phi}_{\tc{b}} }\hat{\rho}_\omega e^{ - i s'_{\tc{a}}  \hat{\phi}_{\tc{a}} - i s'_{\tc{b}}  \hat{\phi}_{\tc{b}} }   \ket{s_{\tc{a}},s_{\tc{b}}}\!\!\bra{s'_{\tc{a}},s'_{\tc{b}}},
\end{align}
where we have set $\bra{s_a}\hat{\rho}_{a,0} \ket{s'_a}=\frac{1}{2}$ for $a=\text{A,B}$. Define the reduced density operator $\hat{\sigma}_\tc{ab} = \tr_\phi(\hat{U}_h \hat{\rho}_0\hat{U}_h^\dagger)$, which corresponds to the partial trace over the field. The remaining terms in the full unitary evolution are independent of the trace over $\phi$. We find
\begin{align}\label{sigma_ab}
    &\hat{\sigma}_\tc{ab} 
    =\frac{1}{4}\sum_{{\substack{{}_{s_{\tc{a}}s_{\tc{b}} }\\{}_{s'_{\tc{a}} s'_{\tc{b}}}}}}  \omega\!\left(e^{ - i s'_{\tc{a}}  \hat{\phi}_{\tc{a}}  - i s'_{\tc{b}}  \hat{\phi}_{\tc{b}} } e^{ i s_{\tc{a}}  \hat{\phi}_{\tc{a}} + i s_{\tc{b}}  \hat{\phi}_{\tc{b}} }\right) \ket{s_{\tc{a}},s_{\tc{b}}}\!\!\bra{s'_{\tc{a}},s'_{\tc{b}}}.
    \end{align}
Using the Weyl relations and that $\omega(e^{i \lambda \hat{\phi}(f)})= e^{- \frac{\lambda^2}{2} W(f,f)}$ for Gaussian states, 
\begin{equation} \label{Gaussian}
    \omega\left(e^{i \lambda \hat{\phi}(f)}e^{i \lambda \hat{\phi}(g)}\right) = e^{-\frac{i \lambda^2}{2}E(f,g) - \frac{\lambda^2}{2} W(f+g,f+g)},
\end{equation}
where $i E(f,g)=[\hat{\phi}(f), \hat{\phi}(g)]$ the commutator and $W(f,g)= \omega ( \hat{\phi}(f)\hat{\phi}(g) )$ the two-point function. Define the Hadamard distribution
\begin{equation}
\label{eq:HadamardDef}
    H(f,g) = \omega\big(\{\hat{\phi}(f),\hat{\phi}(g)\}\big),
\end{equation}
so that the decomposition \mbox{$\hat{\phi}(f)\hat{\phi}(g) = \frac{1}{2}\{\hat{\phi}(f),\hat{\phi}(g)\}+\frac{1}{2}[\hat{\phi}(f),\hat{\phi}(g)]$} allows us to decompose the Wightman as
\begin{align}\label{eq:WHE}
    W(f,g) &= \frac{1}{2}H(f,g) + \frac{i}{2}E(f,g),
\end{align}
where $E(f,g)$ is the causal propagator. We note that the commutator $[\hat{\phi}(f),\hat{\phi}(g)] = E(f,g)\openone$ is proportional to the identity operator, so that it is state independent, as $\omega(\openone) = 1$ for any state. Thus, the Hadamard function contains all the state dependence of the field's correlations. Further, note that in~\eqref{Gaussian} the Wightman function reduces to the Hadamard function, which is symmetric in its arguments, since the antisymmetric component, the causal propagator, is zero when evaluated on $(f+g,f+g)$.

The final step is to include the effect of~\eqref{U_c} on the reduced state of the masses. The full reduced density matrix of A and B is then given by
\begin{equation}
    \hat{\rho}^f_\tc{ab} = \hat{U}_\tc{c} \hat{\sigma}_\tc{ab}\hat{U}_\tc{c}^\dagger.\label{eq:sab}
\end{equation} 
Notice that the operator $\hat{U}_\tc{c}$ is diagonal in the basis spanned by $ |s_a, s_b\rangle$. Using \eqref{sigma_ab} and \eqref{eq:sab}
\begin{equation}
     \hat{\rho}^f_\tc{ab} =\frac{1}{4}\sum_{{\substack{{}_{s_{\tc{a}}s_{\tc{b}} }\\{}_{s'_{\tc{a}} s'_{\tc{b}}}}}}  \omega\!\left(e^{ - i s'_{\tc{a}}  \hat{\phi}_{\tc{a}} - i s'_{\tc{b}}  \hat{\phi}_{\tc{b}} } e^{ i s_{\tc{a}}  \hat{\phi}_{\tc{a}} + i s_{\tc{b}}  \hat{\phi}_{\tc{b}} }\right) e^{-\frac{i}{2}\Delta_{\tc{ab}} (s_{\tc{a}}s_{\tc{b}}- s'_{\tc{a}}s'_{\tc{b}})}\ket{s_{\tc{a}},s_{\tc{b}}}\!\!\bra{s'_{\tc{a}},s'_{\tc{b}}},
\end{equation}
where the self interaction terms have cancelled out. We can then write the explicit form of the time evolved density operator $\hat{\rho}_{\tc{ab}}$, by evaluating the coefficients $\omega(...)$ using \eqref{Gaussian}. It reads: 
\begin{equation}
    \hat{\rho}^f_\tc{ab} = \frac{1}{4}\begin{pmatrix}
        1 & e^{-{\gamma}_{\tc{b}} - i(E_\tc{ab} - \Delta_{\tc{ab}})} & e^{-{\gamma}_{\tc{a}} + i(E_\tc{ab} + \Delta_{\tc{ab}})} & e^{-(\gamma_{\tc{a}}+\gamma_{\tc{b}}+2\gamma_{\tc{ab}})} \\
        e^{-\gamma_{\tc{b}} + i(E_\tc{ab} - \Delta_{\tc{ab}})} & 1 & e^{-(\gamma_{\tc{a}}+\gamma_{\tc{b}}-2\gamma_{\tc{ab}})} & e^{-\gamma_{\tc{a}} - i(E_\tc{ab} + \Delta_{\tc{ab}})} \\
        e^{-\gamma_{\tc{a}} - i(E_\tc{ab} + \Delta_{\tc{ab}})} & e^{-(\gamma_{\tc{a}}+\gamma_{\tc{b}}-2\gamma_{\tc{ab}})} & 1 & e^{-\gamma_{\tc{b}} + i(E_\tc{ab} - \Delta_{\tc{ab}})} \\
        e^{-(\gamma_{\tc{a}}+\gamma_{\tc{b}}+2\gamma_{\tc{ab}})} & e^{-\gamma_{\tc{a}} + i(E_\tc{ab} + \Delta_{\tc{ab}})} & e^{-\gamma_{\tc{b}} - i(E_\tc{ab} - \Delta_{\tc{ab}})} & 1 
    \end{pmatrix}.
\end{equation}
where the $\gamma$ terms are given by convolutions of the Hadamard function
\begin{align}
    \gamma_{\tc a \tc b}& =\iint \dd t\,\dd t'\;\lambda_\tc{a}(t)H(t,\boldsymbol{x}_a\,;t',\bm x_b)\,\lambda_\tc{b}(t') ,\nonumber \\
\gamma_{a}& =\iint \dd t\,\dd t'\;\lambda_a(t)H(t,\boldsymbol{x}_a\,;t',\bm x_a)\,\lambda_a(t'),
\end{align}
for $a=\text{A,B}$. Using that the symmetric and the causal propagators are combined as follows
\begin{align}
\label{eq:bigGammaFull}
    \Delta_{\tc{a}\tc{b}} - E_{\tc{a}\tc{b}} &= 2 G^{\tc{ba}}_{\tc{r}}\equiv 2\varphi_\tc{a}^\tc{b},\\
    \Delta_{\tc{a}\tc{b}} + E_{\tc{a}\tc{b}} &= 2G^{{\tc{ab}}}_{\tc{r}}\equiv 2\varphi_\tc{b}^\tc{a},
\end{align}
then one gets the reduced density matrix \eqref{eq:toyrhofFull}.

\subsection*{Path superposition model}

The solution of the path-superposition model dynamics proceeds analogously. The interaction Hamiltonian~\eqref{eq:HAHB} corresponds to the density
\begin{equation}
    \hat{h}_{int}(\mf x)= \hat{P}^\tc{a}_{\tc{l}} \hat{\phi}^\tc{a}_{\tc{l}}(\mf x)+  \hat{P}^\tc{a}_{\tc{r}} \hat{\phi}^\tc{a}_{\tc{r}}(\mf x)+\hat{P}^\tc{b}_{\tc{l}} \hat{\phi}^\tc{b}_{\tc{l}}(\mf x)+  \hat{P}^\tc{b}_{\tc{r}} \hat{\phi}^\tc{b}_{\tc{r}}(\mf x)
\end{equation}
where $\hat{P}^a_{s}=\ket{s}\bra{s}_a$, $s=\text{L,R}$, $a=\text{A,B}$ and $\hat{\phi}^a_{s}(\mf x)=\lambda_a(t)\delta^3(\bm x-\bm x_s^a(t))\hat{\phi}(\mf x)$ the field operator evaluated over the $s=\text{L,R}$ trajectory of mass $a=\text{A,B}$. We can define $\Lambda_{as}(\mf x)=\lambda_a(t)\delta^3(\bm x-\bm x_s^a(t))$ the smearing of the field operator. Note that, compared to the toy-model, we now explicitly model the path superposition through $\bm x_s^a(t)$.

Since it holds that $[[\hat{h}(\mf x),\hat{h}(\mf x')],\hat{h}(\mf x'')]=  0$ we again use the Magnus expansion~\eqref{eq:AppMagnus}, where
\begin{align}
    \hat{\Theta}_1 &= - i \int \dd V \hat{h}_{int}(\mf x) =  -i\sum_{{\substack{{}_{a=\tc{a},\tc{b} }\\{}_{s=\tc{l},\tc{r}}}}} \hat{P}^a_{s}\hat{\phi}^a_{s},\\
    \hat{\Theta}_2 &= - \frac{1}{2} \int \dd V \dd V' \theta(t-t')[\hat{h}_{int}(\mf x), \hat{h}_{int}(\mf x')] \nonumber\\
    &= \frac{i}{2}\sum_{{\substack{{}_{a,b=\tc{a},\tc{b} }\\{}_{s_{{a}},s_{{b}}=\tc{l},\tc{r}}}}} \varphi^{as_a}_{bs_b}\hat{P}^a_{s_a}\hat{P}^b_{s_b},
\end{align}
where $ \hat{\phi}^a_s= i \int \dd V  \hat{\phi}^{a}_s (\mathsf{x})$ for $a=\text{A,B}$, $s=\text{L,R}$. The phases $\varphi^{as_a}_{bs_b}$ are convolutions of the retarded Green's function, as defined in Section~\ref{sec:pathModel}
\begin{equation}
    \varphi^{{as_a}}_{bs_b}=\iint \dd t\,\dd t'\;\lambda_{a}(t)G_{\tc r}(t, \bm x_{s_a}^{{a}}(t)\,;t', \bm x_{s_b}^{{b}} (t'))\lambda_a(t').
\end{equation}
Note that terms corresponding to the same mass but different paths drop out due to the projectors $\hat{P}^a_s\hat{P}^a_{s'}=\delta_{ss'}\hat{P}^a_s$. Self interaction terms $\varphi^{as}_{as}$ will also drop out from the expression of the reduced states.

Due to $[\hat{\Theta}_1,\hat{\Theta}_2]=0$ we decompose the unitary time evolution operator
\begin{equation}
    \hat{U}_{int} =e^{\hat{\Theta}_1}e^{\hat{\Theta}_2}\equiv \hat{U}_\phi\hat{U}_\tc{c}
\end{equation}
and proceed as we did for the toy-model to find
\begin{equation}
     \hat{\rho}^f_\tc{ab} =\frac{1}{4}\sum_{{\substack{{}_{s_{\tc{a}}s_{\tc{b}} }\\{}_{s'_{\tc{a}} s'_{\tc{b}}}}}}  \omega\!\left(e^{  i  \hat{\phi}^{\tc{a}}_{s'_\tc{a}} + i \hat{\phi}^{\tc{b}}_{s'_\tc{b}} } e^{ -i \hat{\phi}^{\tc{a}}_{s_\tc{a}} - i \hat{\phi}^{\tc{b}}_{s_\tc{b}} }\right) e^{\frac{i}{2} (\varphi^{\tc{a}s_\tc{a}}_{\tc{b}s_\tc{b}}+\varphi_{\tc{a}s_\tc{a}}^{\tc{b}s_\tc{b}} -\varphi^{\tc{a}s'_\tc{a}}_{\tc{b}s'_\tc{b}} -\varphi_{\tc{a}s'_\tc{a}}^{\tc{b}s'_\tc{b}})}\ket{s_{\tc{a}},s_{\tc{b}}}\!\!\bra{s'_{\tc{a}},s'_{\tc{b}}}.
\end{equation}
Using the Weyl relation~\eqref{Gaussian} we expand the coefficients
\begin{equation}
    \begin{aligned}
       \omega\!\left(e^{  i  \hat{\phi}^{\tc{a}}_{s'_\tc{a}} + i \hat{\phi}^{\tc{b}}_{s'_\tc{b}} } e^{ -i \hat{\phi}^{\tc{a}}_{s_\tc{a}} - i \hat{\phi}^{\tc{b}}_{s_\tc{b}} }\right)&=  e^{\frac{i}{2}E(\Lambda^\tc{a}_{s_\tc{a}}+\Lambda^\tc{b}_{s_\tc{b}},\Lambda^\tc{a}_{s'_\tc{a}}+\Lambda^\tc{b}_{s'_\tc{b}}) - \frac{1}{2} W(\Lambda^\tc{a}_{s_\tc{a}}+\Lambda^\tc{b}_{s_\tc{b}}-\Lambda^\tc{a}_{s'_\tc{a}}-\Lambda^\tc{b}_{s'_\tc{b}},\Lambda^\tc{a}_{s_\tc{a}}+\Lambda^\tc{b}_{s_\tc{b}}-\Lambda^\tc{a}_{s'_\tc{a}}-\Lambda^\tc{b}_{s'_\tc{b}})}\\
       &=  e^{\frac{i}{2}(\varphi^{\tc{a}s_\tc{a}}_{\tc{b}s'_\tc{b}}+\varphi^{\tc{b}s_\tc{b}}_{\tc{a}s'_\tc{a}}- \varphi_{\tc{a}s_\tc{a}}^{\tc{b}s'_\tc{b}}-\varphi_{\tc{b}s_\tc{b}}^{\tc{a}s'_\tc{a}}) - \frac{1}{4} H(\Lambda^\tc{a}_{s_\tc{a}}+\Lambda^\tc{b}_{s_\tc{b}}-\Lambda^\tc{a}_{s'_\tc{a}}-\Lambda^\tc{b}_{s'_\tc{b}},\Lambda^\tc{a}_{s_\tc{a}}+\Lambda^\tc{b}_{s_\tc{b}}-\Lambda^\tc{a}_{s'_\tc{a}}-\Lambda^\tc{b}_{s'_\tc{b}})}.
       \label{eq:AppCoef}
    \end{aligned}
\end{equation}
Evaluating the coefficients and phases for all path configurations we derive the final density matrix~\eqref{eq:rhofpath}. In the case of path superposition the fluctuation terms, which are the convolutions of the Hadamard function in~\eqref{eq:AppCoef}, sample both trajectories followed by each mass. Specifically:
\begin{equation}
    \begin{aligned}
        4\Gamma_{\tc{a}} &= \iint \dd t\,\dd t' \lambda_\tc{a}(t) \lambda_\tc{a}(t')\Bigl(H(t,\bm x_{\tc l}^{\tc{a}}(t)\,;t',\bm x_{\tc l}^{\tc{a}}(t')) + H(t,\bm x_{\tc r}^{\tc{a}}(t)\,;t',\bm x_{\tc r}^{\tc{a}}(t'))- 2H(t,\bm x_{\tc l}^{\tc{a}}(t)\,;t',\bm x_{\tc r}^{\tc{a}}(t'))  \Bigr)\,, \\ 
        4\Gamma_{\tc{b}} &= \iint \dd t\,\dd t' \lambda_\tc{b}(t) \lambda_\tc{b}(t')\Bigl(H(t,\bm x_{\tc l}^{\tc{b}}(t)\,;t',\bm x_{\tc l}^{\tc{b}}(t')) + H(t,\bm x_{\tc r}^{\tc{b}}(t)\,;t',\bm x_{\tc r}^{\tc{b}}(t'))- 2H(t,\bm x_{\tc l}^{\tc{b}}(t)\,;t',\bm x_{\tc r}^{\tc{b}}(t'))  \Bigr)\,, \\
    4\Gamma_{\tc{ab}} &=  \iint \dd t\,\dd t' \lambda_\tc{a}(t) \lambda_\tc{b}(t')\Bigl(H(t,\bm x_{\tc l}^{\tc{a}}(t)\,;t',\bm x_{\tc l}^{\tc{b}}(t')) + H(t,\bm x_{\tc r}^{\tc{a}}(t)\,;t',\bm x_{\tc r}^{\tc{b}}(t')) \\ &\quad\quad\quad\quad\quad\quad\quad\quad\quad\quad\quad-H(t,\bm x_{\tc l}^{\tc{a}}(t)\,;t',\bm x_{\tc r}^{\tc{b}}(t')) - H(t,\bm x_{\tc r}^{\tc{a}}(t)\,;t',\bm x_{\tc l}^{\tc{b}}(t')) \Bigr)\,.
    \end{aligned}
\end{equation}

In terms of the toy-model definitions~\eqref{eq:toyGHterms}:
\begin{equation}
\begin{aligned}
    4\Gamma_{\tc{a}}&=\gamma_{\tc{a}}^{\tc{ll}} +\gamma_{\tc{a}}^{\tc{rr}} -2 \gamma_{\tc{a}}^{\tc{lr}} \\
    4\Gamma_{\tc{b}}&=\gamma_{\tc{b}}^{\tc{ll}} +\gamma_{\tc{b}}^{\tc{rr}} -2 \gamma_{\tc{b}}^{\tc{lr}} \\
    4\Gamma_{\tc{ab}}&=\gamma_{\tc{ab}}^{\tc{ll}} +\gamma_{\tc{ab}}^{\tc{rr}} - \gamma_{\tc{ab}}^{\tc{lr}} - \gamma_{\tc{ab}}^{\tc{rl}}
\end{aligned}
\end{equation}

\section{Fluctuation terms in the path superposition model}
\label{sec:AppFluct}
In this appendix, we prove that the quantum fluctuation terms in the path-superposition model of Section~\ref{sec:pathModel} lead to decoherence. These terms appear in negative exponentials in the off-diagonal elements of the final state~\eqref{eq:rhofpath}, so it suffices to show that
\begin{equation}
\begin{aligned}
    \Gamma_{\tc{a}}&\geq 0,\\
    \Gamma_{\tc{b}}&\geq 0,\\
    \Gamma_{\tc{a}}+ \Gamma_{\tc{b}}\pm  2\Gamma_{\tc{ab}}&\geq 0.\\
\end{aligned}
\end{equation}

We extend the proof for the positivity of $\gamma_a$ terms of the toy-model of Section~\ref{sec:Scalar} found in~\cite{2023hidaka}. In short, we express the $\Gamma_a$ terms as norms of quantum states, which are by definition positive. Indeed, if we define the states
\begin{equation}
\begin{aligned}
\ket{A}&=\lambda_{\tc{a}}\int_{t_i}^{t_f} dt \left(\hat{\phi}(x^{\tc a}_\tc{l},t) - \hat{\phi}(x^{\tc a}_\tc{r},t) \right) \ket{\omega}_\phi ,\\
\ket{B}&=\lambda_{\tc{b}}\int_{t_i}^{t_f} dt \left(\hat{\phi}(x^{\tc b}_\tc{l},t) - \hat{\phi}(x^{\tc b}_\tc{r},t) \right) \ket{\omega}_\phi,
\end{aligned}
\end{equation}
we find
\begin{equation}
\begin{aligned}
    \Gamma_{\tc{a}}=2\braket{A}{A}&\geq 0,\\
     \Gamma_{\tc{b}}=2\braket{B}{B}&\geq 0.
\end{aligned}
\end{equation}
For the sum/difference terms we define the states
\begin{equation}
    \ket{A\pm B}= \ket{A}\pm \ket{B}
\end{equation}
to find $\Gamma_{\tc{a}}+ \Gamma_{\tc{b}}\pm  2\Gamma_{\tc{ab}}= 2\braket{A\pm B}{A\pm B}\geq 0$.

\section{Stationary phase approximation in the path superposition model} \label{SPA}
In this appendix, we solve the path integral for the scalar field under a stationary phase approximation and derive equation~\eqref{eq:toyOverlapsOS}. We follow the steps in the supplementary material of~\cite{Christodoulou:2022mkf}, that is we start from a path integral expression of the dynamics and then insert the solution to the classical equations of motion for the field, ignoring all other contributions to the path integral.

We start with the path-integral expression we want to approximate, corresponding to equation~\eqref{eq:toyOverlapsPI} of the main text
\begin{equation}
\label{eq:appPI}
\langle \omega^f_{\sigma'} |\omega^f_{\sigma}\rangle  ={ \int }   \mathcal{D}\left[ \phi(\mf x) \right]\,  \mathcal{D}\left[ \phi'(\mf x) \right]\,  e^{ i  S_{CTP}\left[ \phi,\phi';J_{\sigma},J_{\sigma'} \right] } 
\end{equation}
where the action in the CTP formalism~\eqref{sbmveqSCTP}
\begin{align}
S_{CTP}[\phi,\phi';J_{\sigma},J_{\sigma'}]=\int  \dd^4\mf x\big( -\frac{1}{2}\partial_{\mu}\phi\partial^\mu \phi+\phi J_{\sigma} + \frac{1}{2}\partial_{\mu}\phi'\partial^\mu \phi'-\phi'J_{\sigma'} \big).
\end{align}
We remind that the field satisfies vacuum boundary conditions for $t=t_i$, while for $t=t_f$ $\phi(t_{f}, \boldsymbol{x})=\phi'(t_{f}, \boldsymbol{x})$. Next, we calculate the classical field configurations by solving the classical equations of motion. For the scalar field, these take the form of a Klein-Gordon equation
    \begin{align}
 \square \phi_{\text{cl}}(x)&=-J_{\sigma}(x)\nonumber \\
  \square \phi'_\text{cl}(x)&=-J_{\sigma'}(x)
\end{align}
where $\square=\partial_\mu\partial^\mu$ and our convention for the metric signature is $(-,+,+,+)$. Due to the vacuum initial conditions, the solution is given in terms of the retarded Green's function
\begin{align}
\label{eq:appclass}
    \phi_{\text{cl}}(\mf x) &= \int \dd^4 \mf y \, G_{\tc{r}}(\mf x ; \mf y) J_{\sigma}(\mf y)\nonumber\\
    \phi'_{\text{cl}}(\mf x)  &= \int \dd^4 \mf y \, G_{\tc{r}}(\mf x ; \mf y) J_{\sigma'}(\mf y)
\end{align}
We integrate the action by parts
\begin{equation}
    S_{CTP}[\phi,\phi';J_{\sigma},J_{\sigma'}]= \int \dd^4\mf{x}\left( \frac{1}{2}\phi_{}\square\phi_{}+J_{\sigma}\,\phi_{}- \frac{1}{2}\phi'_{}\square\phi'_{}-J_{\sigma'}\,\phi'_{}\right).
\end{equation}
Then, the action calculated on the classical field configurations~\eqref{eq:appclass} is
\begin{equation}
     S_{CTP}[\phi_\text{cl},\phi'_\text{cl}  ;J_{\sigma},J_{\sigma'}]= \frac{1}{2} \iint  \dd^4\mf x \dd^4\mf y \Bigl( J_{\sigma}(\mf x)G_{\tc r}(\mf x;\mf y) J_{\sigma}(\mf y) - J_{\sigma'}(\mf x)G_{\tc r}(\mf x;\mf y) J_{\sigma'}(\mf y)\Bigr)
\end{equation}
This quantity has been previously called the on-shell action~\cite{Christodoulou:2022mkf}.

Finally, the stationary phase approximation of the path integral amounts to inserting the on-shell action in~\eqref{eq:appPI} and ignoring all other contributions to the path integral
\begin{equation}
\begin{aligned}
\langle\omega_{\sigma'}^f| \omega_{\sigma}^f \rangle \approx e^{iS_{CTP}[\phi_\text{cl},\phi'_\text{cl}  ;J_{\sigma},J_{\sigma'}]}=\mathrm{exp} \Biggl(\frac{i}{2} \iint  \dd^4\mf x \dd^4\mf y \Bigl( J_{\sigma}(\mf x)G_{\tc r}(\mf x;\mf y) J_{\sigma}(\mf y) - J_{\sigma'}(\mf x)G_{\tc r}(\mf x;\mf y) J_{\sigma'}(\mf y)\Bigr) \Biggr)
\end{aligned}
\end{equation}
This is equation~\eqref{eq:toyOverlapsOS} of the main text, from which one can calculate the reduced state of the masses under the stationary phase approximation.

\section{Robertson-Schrödinger inequality in the path superposition model}
\label{sec:AppUnc}
In this appendix, we derive the Robertson-Schrödinger inequality, which is equation~\eqref{eq:RSineq} of Section~\ref{sec:NoPertinenceNewton}. In short, we use the uncertainty principle for field operators,  generalizing to the path-superposition model the proof found  in~\cite{2023hidaka}.

Consider the field operators
\begin{equation}
    \begin{aligned}
  \hat{\mathcal{O}}_\tc{a} &=\lambda_{\tc a}\int_{t_i}^{t_f} \dd t \left(\hat{\phi}(x^{\tc a}_\tc{l},t) - \hat{\phi}(x^{\tc a}_\tc{r},t) \right),\\
\hat{\mathcal{O}}_\tc{b} &=\lambda_{\tc b}\int_{t_i}^{t_f}\dd t \left(\hat{\phi}(x^{\tc b}_\tc{l},t) - \hat{\phi}(x^{\tc b}_\tc{r},t) \right).
    \end{aligned}
    \label{eq:AppOps}
\end{equation}
These operators are Hermitian and have vanishing expectation values. The Robertson-Schrödinger inequality for two such operators is~\cite{2023hidaka}
\begin{equation}
    \langle \hat{\mathcal{O}}_\tc{a}^2\rangle \langle \hat{\mathcal{O}}_\tc{b}^2\rangle \geq \frac{1}{4} \left( |\langle [\hat{\mathcal{O}}_\tc{a},\hat{\mathcal{O}}_\tc{b}]\rangle|^2 + |\langle \{\hat{\mathcal{O}}_\tc{a},\hat{\mathcal{O}}_\tc{b}\}\rangle|^2 \right).
\end{equation}
For the specific operators~\eqref{eq:AppOps} we find
\begin{equation}
\begin{aligned}
    \langle \hat{\mathcal{O}}_{a}^2\rangle &=\frac{1}{2}\Gamma_{aa}\,,\;a=A,B\\
    \langle \{\hat{\mathcal{O}}_\tc{a},\hat{\mathcal{O}}_\tc{b}\}\rangle &= \Gamma_{\tc{ab}}\\
    \langle [\hat{\mathcal{O}}_\tc{a},\hat{\mathcal{O}}_\tc{b}]\rangle &= (\varphi^{\tc b}_{\tc{ar}}-\varphi^{\tc a}_{\tc{br}}) - (\varphi^{\tc b}_{\tc{al}}-\varphi^{\tc a}_{\tc{bl}})
\end{aligned}
\end{equation}
and the Robertson-Schrödinger inequality becomes
\begin{equation}
\label{eq:RSapp}
\Gamma_{\tc{a}}\Gamma_{\tc{b}}-\Gamma^2_{\tc{ab}}\geq\left((\varphi^{\tc b}_{\tc{ar}}-\varphi^{\tc a}_{\tc{br}}) - (\varphi^{\tc b}_{\tc{al}}-\varphi^{\tc a}_{\tc{bl}})\right)^2
\end{equation}
In Section~\ref{sec:Paradox} of the main text, we make use of the weaker inequality~\eqref{eq:RSineq}
\begin{equation}
\Gamma_{\tc{a}}\Gamma_{\tc{b}}\geq\left((\varphi^{\tc b}_{\tc{ar}}-\varphi^{\tc a}_{\tc{br}}) - (\varphi^{\tc b}_{\tc{al}}-\varphi^{\tc a}_{\tc{bl}})\right)^2.
\end{equation}
which is implied by \eqref{eq:RSapp}.


\section{Visibility/distinguishability calculations}
\label{sec:AppVisDis}
\label{sec:AppInfo}

In this appendix, we calculate the visibility of A and the distinguishability of B, for the approximations
 relevant to the discussion of the paradox in Section~\ref{sec:Paradox}.  We fix the set-up depicted in Figure~\ref{fig:paradox}, for which $\varphi^{\tc{a}}_{\tc{b}s_\tc{b}}= 0$.

\subsection*{Visibility of A}

The definition of the visibility of A is $\mathcal{V}_\tc{a}=2|\bra{L} \hat{\rho}_\tc{a}(t_f)\ket{R}_\tc{a}|$, which is calculated directly from the off-diagonal element of the state of A. We consider the following cases.

\smallskip
{\bf (i). No approximation}

Before ignoring the quantum fluctuations, the reduced state of A is given in the main text as~\eqref{eq:reducedA}
\begin{equation}
\hat{\rho}_\tc{a}(t_f)= 
\frac{1}{2}\begin{pmatrix}
1  &  e^{-\Gamma_{\tc{a}}} \\ 
 e^{-\Gamma_{\tc{a}}}   & 1
\end{pmatrix} 
\end{equation}
from which $\mathcal{V}_\tc{a}=e^{-\Gamma_{\tc{a}}}$. That is, the visibility only contains the local fluctuation term $\Gamma_\tc{a}$.

\smallskip
{\bf (ii). Setting $\mathbf{\Gamma=0}$}

The reduced state of A when setting $\Gamma=0$ is the coherent superposition
\begin{equation}
\hat{\rho}_\tc{a}(t_f)= 
\frac{1}{2}\begin{pmatrix}
1  &  1 \\ 
1   & 1
\end{pmatrix} 
\end{equation}
so $\mathcal{V}_\tc{a}=1$. In this case, A  does not decohere.

\smallskip
{\bf (iii). Stationary phase approximation}
    
The state of A after a stationary phase approximation~\eqref{eq:SPAreducedA}
\begin{equation}
\hat{\rho}_\tc{a}(t_f)= 
\frac{1}{2}\begin{pmatrix}
\;\;1  &  \frac{1}{2}(e^{ \frac{i}{2} \Delta_{\tc a}^{\tc{b}\tc{l}}} + e^{  \frac{i}{2}  \Delta_{\tc a}^{\tc{b}\tc{r}} } ) \\ 
 & 1
\end{pmatrix} ,
\end{equation}
from which
\begin{equation}
\begin{aligned}
    \mathcal{V}_\tc{a}&=\frac{1}{2}\left| e^{ \frac{i}{2} \Delta_{\tc a}^{\tc{b}\tc{l}} } + e^{  \frac{i}{2}  \Delta_{\tc a}^{\tc{b}\tc{r}} }\right|\\
    &=\frac{1}{2} \sqrt{\left(\cos{\frac{\Delta_{\tc a}^{\tc{b}\tc{l}}}{2}} +   \cos{\frac{\Delta_{\tc a}^{\tc{b}\tc{r}}}{2}} \right)^2+ \left(\sin{\frac{\Delta_{\tc a}^{\tc{b}\tc{l}}}{2}}  +  \sin{\frac{\Delta_{\tc a}^{\tc{b}\tc{r}}}{2}}\right)^2}\\
    &=\frac{1}{2} \sqrt{2+2 \cos{\frac{\Delta_{\tc a}^{\tc{b}\tc{l}} - \Delta_{\tc a}^{\tc{b}\tc{r}}}{2}} }\\
    &=\left| \cos\left(\frac{\varphi^{\tc b}_{\tc{al}}-\varphi^{\tc b}_{\tc{ar}}}{4} \right)\right|
\end{aligned}
\end{equation}
which is equation~\eqref{eq:SPAVD} of the main text.

\subsection*{Distinguishability of B}

The definition of the distinguishability of B was given in Section~\ref{sec:Paradox} as $\mathcal{D}_{\tc b}=\frac{1}{2} \mid\mid \hat{\rho}^{\tc l}_{\tc b}-\hat{\rho}^{\tc r}_{\tc b} \mid\mid_{1}$, where $\mid\mid \cdot \mid\mid_{1}$ is the trace norm, and $\hat{\rho}^{\tc{l}/\tc{r}}_{\tc b}$ is the state of B when A follows the left/right path:
$$
\begin{aligned}
\hat{\rho}^{\tc l}_{\tc b}&=2\,\langle L |\hat{\rho}_{\tc{ab}}  | L \rangle_{\tc a} \\
\hat{\rho}^{\tc r}_{\tc b}&=2\,\langle R |\hat{\rho}_{\tc{ab}}  | R \rangle_{\tc a} 
\end{aligned}
$$ 
A simple method to calculate the distinguishability is to express $\hat{\rho}^{\tc l}_{\tc b}$ and $\hat{\rho}^{\tc r}_{\tc b}$ as vectors in the Bloch representation. Then, the trace distance is given by half the Euclidean distance of the two vectors in the Bloch representation.~\cite{Hidaka:2022gsv}

\smallskip
{\bf (i). No approximation}

Including quantum fluctuations, we read $\hat{\rho}^{\tc l}_{\tc b}$ and $\hat{\rho}^{\tc r}_{\tc b}$ from the total state~\eqref{eq:rhofpath}

    \begin{align}
    \label{eq:apprhol}
\hat{\rho}^{\tc{l}}_\tc{b}(t_f)= 
\frac{1}{2}\begin{pmatrix}
1  &  e^{-\Gamma_{\tc{b}}}  e^{ i \varphi^{\tc b}_{\tc{al}}}\\ 
e^{-\Gamma_{\tc{b}}}  e^{ -i \varphi^{\tc b}_{\tc{al}}} & 1 \end{pmatrix} =\frac{1}{2} \left(\hat{\openone}+ e^{-\Gamma_{\tc{b}}} \cos{(\varphi^{\tc b}_{\tc{al}})}\hat{\sigma}_\text{x}- e^{-\Gamma_{\tc{b}}}\sin{(\varphi^{\tc b}_{\tc{al}})}\hat{\sigma}_\text{y} \right),\\
\label{eq:apprhor}
\hat{\rho}^{\tc{r}}_\tc{b}(t_f)= 
\frac{1}{2}\begin{pmatrix}
1  &  e^{-\Gamma_{\tc{b}}}  e^{ i \varphi^{\tc b}_{\tc{ar}}}\\ 
e^{-\Gamma_{\tc{b}}}  e^{ -i \varphi^{\tc b}_{\tc{ar}}} & 1 \end{pmatrix}=\frac{1}{2} \left(\hat{\openone}+ e^{-\Gamma_{\tc{b}}}\cos{(\varphi^{\tc b}_{\tc{ar}})}\hat{\sigma}_\text{x}- e^{-\Gamma_{\tc{b}}} \sin{(\varphi^{\tc b}_{\tc{ar}})}\hat{\sigma}_\text{y} \right).
    \end{align}
 
Then, the distinguishability is
\begin{equation}
\label{eq:D_Bfull}
\begin{aligned}
    \mathcal{D}_{\tc b}&= \frac{e^{-\Gamma_\tc{b}}}{2} \sqrt{\left(\cos{ \varphi^{\tc b}_{\tc{al}} } -   \cos{\varphi^{\tc b}_{\tc{ar}}} \right)^2+ \left(\sin{\varphi^{\tc b}_{\tc{al}}}  -  \sin{\varphi^{\tc b}_{\tc{ar}}}\right)^2}\\
    &=e^{-\Gamma_\tc{b}}\left| \sin{\frac{(\varphi^{\tc b}_{\tc{al}}-\varphi^{\tc b}_{\tc{ar}})}{2}} \right|\\
\end{aligned}
\end{equation}
From this we also see that the `decoherence-inducing radiation' produced by A $(\Gamma_{\tc{a}})$ does not enter into the which path information available to B.

\smallskip
{\bf (ii). Setting $\mathbf{\Gamma=0}$}

We directly compute this case by setting $\Gamma_\tc{b}=0$ in~\eqref{eq:D_Bfull}
\begin{equation}
    \mathcal{D}_{\tc b}=\left| \sin{\frac{(\varphi^{\tc b}_{\tc{al}}-\varphi^{\tc b}_{\tc{ar}})}{2}} \right|
\end{equation}
which is equation~\eqref{eq:DBGamma=0} of the main text.

\smallskip
{\bf (iii). Stationary phase approximation}

For the stationary phase approximation~\eqref{eq:rhofSP}
 \begin{align}
\hat{\rho}^{\tc{l}}_\tc{b}(t_f)= 
\frac{1}{2}\begin{pmatrix}
1  &   e^{ \frac{i}{2} \varphi^{\tc b}_{\tc{al}}}\\ 
  e^{ -\frac{i}{2} \varphi^{\tc b}_{\tc{al}}} & 1 \end{pmatrix} =\frac{1}{2} \left(\hat{\openone}+  \cos{\left(\frac{\varphi^{\tc b}_{\tc{al}}}{2}\right)}\hat{\sigma}_\text{x}- \sin{\left(\frac{\varphi^{\tc b}_{\tc{al}}}{2}\right)}\hat{\sigma}_\text{y} \right),\\
\hat{\rho}^{\tc{r}}_\tc{b}(t_f)= 
\frac{1}{2}\begin{pmatrix}
1  &    e^{ \frac{i}{2} \varphi^{\tc b}_{\tc{ar}}}\\ 
  e^{ -\frac{i}{2} \varphi^{\tc b}_{\tc{ar}}} & 1 \end{pmatrix}=\frac{1}{2} \left(\hat{\openone}+ \cos{\left(\frac{\varphi^{\tc b}_{\tc{ar}}}{2}\right)}\hat{\sigma}_\text{x}-  \sin{\left(\frac{\varphi^{\tc b}_{\tc{ar}}}{2}\right)}\hat{\sigma}_\text{y} \right).
    \end{align}
By inspection, these are the states~\eqref{eq:apprhol} and~\eqref{eq:apprhor} after setting $\Gamma_\tc{b}=0$ and dividing all the phases by $2$. So, for the stationary phase approximation
\begin{equation}
    \mathcal{D}_{\tc b}=\left| \sin{\frac{(\varphi^{\tc b}_{\tc{al}}-\varphi^{\tc b}_{\tc{ar}})}{4}} \right|
\end{equation}
which is equation~\eqref{eq:SPAVD} of the main text.


\section{No spacelike influences} \label{appendix: reduced states}

In this appendix, we prove that under two assumptions for the dynamics of A, B and the field, there can be no `spacelike influences' between A and B for the paradox setup discussed in Section~\ref{sec:Paradox}. This compels  two conclusions: (a) the state of A does not depend on the choice of agent B to couple, or not, to the field, and (b) the state of B cannot be affected by the recombination process of A, so agent B is not measuring radiation produced by A during recombination.

We start by assuming that the interactions of A and B with the field can be described by local unitaries that transform covariantly and satisfy certain factorisation properties:
\begin{enumerate}
    \item If $\text{B}\in M/J^-(\text{A})$, i.e. B does not intersect the causal past of A (as in the thought experiment) then the interactions are causally orderable, i.e. there is a foliation such that B is `after' A, and it holds that 
    \begin{equation}
        U_{\textsc{a}+\textsc{b}}= U_{\textsc{b}} U_{\textsc{a}}. \label{causal}
    \end{equation}
    \item We can decompose A's interaction w.r.t to any hypersurface $\Sigma$ ($\Sigma_2$ in the thought experiment) that passes through her interaction region as
    \begin{equation}
        U_{\textsc{a}}= U^+_{\textsc{a}} U^{-}_{\textsc{a}} \label{continuous}
  \end{equation}
where $U^{-}$ corresponds to the splitting process (in the causal past of $\Sigma$) and $U^{+}$ corresponds to the recombination process (in the causal future of $\Sigma$)  
\end{enumerate}

Then the total interaction of A and B with the field can be decomposed as
\begin{equation}
     U_{\textsc{a}+\textsc{b}}= U_{\textsc{b}} U^+_{\textsc{a}} U^{-}_{\textsc{a}}. \label{composition}
\end{equation}
It also holds that 
\begin{equation}
    [U^+_{\textsc{a}}, U_{\textsc{b}}]=0 \label{spacelike}
\end{equation}
since in the thought experiment the recombination process of Alice happens in spacelike separation from Bob. Now taking partial traces of the total state after the evolution and using \eqref{composition} and \eqref{spacelike} we can see that:
\begin{enumerate}
    \item The reduced state of A cannot be affected by the existence of B: if $\rho_0$ the initial (uncorrelated) state of A, B and the field, then 
    \begin{equation}\label{NOretrocausality}
        \rho_{\textsc{a}}= \text{tr}_{\textsc{b}\textsc{f}} \left( U_{\textsc{b}} U_{\textsc{a}} \rho_0 U^{\dagger}_{\textsc{a}}  U^{\dagger}_{\textsc{b}}\right)= \text{tr}_{\textsc{b} \textsc{f}} \left(  U^{\dagger}_{\textsc{b}}U_{\textsc{b}} U_{\textsc{a}} \rho_0 U^{\dagger}_{\textsc{a}}  \right) = \text{tr}_{\textsc{f}} \left(  U_{\textsc{a}} \rho_{\textsc{af}} U^{\dagger}_{\textsc{a}}  \right),
    \end{equation} 
    using the cyclic property of the partial trace and the fact that $U^{\dagger}_{\textsc{b}}U_{\textsc{b}}= \textsf{1}_{\textsc{b}}$. Note that we have also used that B should be initially uncorrelated with A and the field, i.e. $\rho_0= \rho_{\textsc{af}} \rho_{\textsc{b}}$, and that $\text{tr}_{\tc{b}} \hat{\rho}_{\tc{b}}=1$. 
    
    We see that the state of B and its interaction with the field drops out of the expression of the A's reduced state, so B can be viewed as an `innocent bystander' (to use the terminology of \cite{2018belenchiaQuantumSuperpositionMassive, 2022danielsonGravitationallyMediatedEntanglement}). Of course, this is as long as $\text{B}\in M/J^-(\text{A})$, which will not hold if A does not follow the `protocol' and recombines slowly. 
    \item The reduced state of B cannot be affected by A's recombination process:
\begin{align}
     \rho_{\textsc{b}}&= \text{tr}_{\textsc{a}\textsc{f}} \left( U_{\textsc{b}} U^+_{\textsc{a}} U^{-}_{\textsc{a}} \rho_0 U^{-\dagger}_{\textsc{a}} U^{+\dagger}_{\textsc{a}}  U^{\dagger}_{\textsc{b}}\right) \nonumber \\
     &= \text{tr}_{\textsc{a}\textsc{f}} \left(  U^{+\dagger}_{\textsc{a}}U^+_{\textsc{a}}U_{\textsc{b}}  U^{-}_{\textsc{a}} \rho_0 U_{\textsc{b}} U^{-\dagger}_{\textsc{a}}   U^{\dagger}_{\textsc{b}}\right) \nonumber \\
     &= \text{tr}_{\textsc{a}\textsc{f}} \left(  U_{\textsc{b}}  U^{-}_{\textsc{a}} \rho_0  U^{-\dagger}_{\textsc{a}}   U^{\dagger}_{\textsc{b}}\right).
\end{align}
We see that the part of the evolution that corresponds to the recombination process $ U^+_{\textsc{a}}$ drops out of the expression of B's reduced state. Note that this holds independently from the shape of the total initial state $\rho_0$, e.g. if it encodes prior correlations between the two parties and the field (or in refined descriptions of the gravitational set-up, initial correlations between the field and B's trap, see also \cite{sudarsky}).
\end{enumerate}


\subsection*{No superluminal signalling for the toy model}

Note that the assumption that \eqref{causal} and \eqref{continuous} hold for the quantum gravitational field (treated as a spin-2 quantum field) is non-trivial. Overall, the derivations above rely on general properties of the local scattering maps that are expected to hold for reasonable interactions/setups. The setup and interactions that are involved in our analysis are simple enough to see explicitly how \eqref{NOretrocausality} holds. To see this (first for the toy model of Section~\ref{sec:Scalar}) using the Magnus expansion (Appendix~\ref{appendix: Magnus}), the total unitary decomposes as
\begin{equation}
    \hat{U}_{\tc{abf}}= e^{- i \mathcal{G}_{\tc{a}}- i \mathcal{G}_{\tc{b}}} e^{- \frac{i}{2} \Delta_{\tc{a}\tc{b}}\hat{\sigma}_\tc{a}\hat{\sigma}_\tc{b}} e^{- i  \hat{\phi}_{\tc{a}} \hat{\sigma}_\tc{a} - i  \hat{\phi}_{\tc{b}} \hat{\sigma}_\tc{b}}
\end{equation} \label{Magnusforhidaka}
where $\hat{\phi}_{\nu}= \int  \dd t\lambda_{\nu}(t) \hat{\phi}_{\nu}(t, \bm{x}_{\nu})$ (where we could also introduce a spatial smearing of the field operator), with $\nu=\text{A},\text{B}$ and $\mathcal{G}_{\nu}$ are the self-interaction terms, e.g.
\begin{equation}
\mathcal{G}_{\nu}= \int \dd t \dd t' \lambda_{\nu}(t)G_{\textsc{r}}(t, \bm{x}_{\textsc{a}}; t',\bm{x}_{\textsc{a}} ) \lambda_{\nu}(t').
\end{equation}
Note that the term $\Delta_{\tc{a}\tc{b}} \hat{\sigma}_\tc{a}\hat{\sigma}_\tc{b}$ in \eqref{Magnusforhidaka}, is the term that survives the stationary phase approximation. $\Delta_{\tc{a}\tc{b}}$ is given by the symmetrised propagator 
\begin{equation} \label{symmetrised}
   \Delta_{\tc{a}\tc{b}}= \int \dd t \dd t' \lambda_{\tc{a}}(t) \left( G_{\tc{r}}(t, \bm{x}_{\tc{a}}; t', \bm{x}_{\tc{b}})+G_{\tc{a}} (t, \bm{x}_{\tc{a}}; t', \bm{x}_{\tc{b}})\right) \lambda_{\tc{b}}(t')
\end{equation}
Once we fix the set-up of the thought experiment, see Section~\ref{sec:Paradox}, in which the retarded phases from Bob to Alice disappear 
\begin{equation} 
    G^{\textsc{ba}}_{\textsc{r}}= \int \dd t \dd t' \lambda_{\tc{a}}(t) G_{\tc{r}}(t, \bm{x}_{\tc{a}}; t', \bm{x}_{\tc{b}}) \lambda_{\tc{b}}(t')=0
\end{equation}
and only the contribution that depends on the advanced propagator survives in \eqref{symmetrised}. Since $G^{\textsc{ba}}_{\textsc{r}}=0$ then  \eqref{symmetrised} is equivalent with the antisymmetrised propagator $E_{\tc{ab}}=G^{\textsc{ba}}_{\textsc{r}}- G^{\textsc{ba}}_{\textsc{a}}$ (which is proportional to the commutator). Then, using the Weyl relations, the total unitary \eqref{Magnusforhidaka} (without any approximations) becomes
\begin{equation}
       \hat{U}_{\tc{abf}}= e^{- i \mathcal{G}_{\tc{a}}- i \mathcal{G}_{\tc{b}}} e^{- i  \hat{\phi}_{\tc{a}} \hat{\sigma}_\tc{a}} e^{ - i  \hat{\phi}_{\tc{b}} \hat{\sigma}_\tc{b}}
\end{equation}
Then, it is straightforward to check that 
\begin{equation}
    \hat{\rho}_{\tc{a}}= \text{tr}_{\tc{bf}} \left( \hat{U}_{\tc{abf}}\hat{\rho}_0 \hat{U}^{\dagger}_{\tc{abf}} \right)= \text{tr}_{\tc{f}} \left( e^{- i  \hat{\phi}_{\tc{a}} \hat{\sigma}_\tc{a}} \hat{\rho}_{\tc{af}} e^{ i  \hat{\phi}_{\tc{a}} \hat{\sigma}_\tc{a}}\right),
\end{equation}
where the quantities that depend on Bob have dropped out, so there is no superluminal signaling.

\section{Newtonian limit}
\label{sec:NewtonianLimit}
For completeness, we briefly discuss here the Newtonian limit. 
Physically, this corresponds to the limit of large interactions times $T\gg D/c$ and small coupling $\lambda\ll 1$. In this limit, we have that $\varphi^{\tc{a}s_\tc{a}}_{\tc{b}s_\tc{b}}=\varphi_{\tc{a}s_\tc{a}}^{\tc{b}s_\tc{b}}$, that is, phases are no longer created with retardation but through `action at a distance'. In this limit, we also have that $\Gamma_a \to 0$, and due to the inequality $\Gamma_{\tc a \tc b}\leq \frac{1}{2}(\Gamma_{\tc a }+\Gamma_{\tc b })$ (see Appendix~\ref{sec:AppFluct} for a derivation), we also have $\Gamma_{\tc a \tc b}\to 0$. As remarked just above, there is no violation of quantum theory in this case, it essentially corresponds to the quantum mechanics of two particles coupled directly by a potential. 

The Newtonian limit of the final state \eqref{eq:rhofpath} is
\begin{equation}
\label{rhofBMVfull}
    \hat{\rho}_\tc{ab}(t_{f})=\frac{1}{4}\left( \begin{matrix}
\;\;1\quad\; &  e^{i(\varphi_{\tc{l}\tc{l}}-\varphi_{\tc{l}\tc{r}})} & e^{i(\varphi_{\tc{l}\tc{l}}-\varphi_{\tc{r}\tc{l}})} & e^{i(\varphi_{\tc{l}\tc{l}}-\varphi_{\tc{r}\tc{r}})}\\
 & 1 & e^{i(\varphi_{\tc{l}\tc{r}}-\varphi_{\tc{r}\tc{l}})}  &  e^{-i(\varphi_{\tc{r}\tc{r}}-\varphi_{\tc{l}\tc{r}})} \\
 &   & 1 & e^{-i(\varphi_{\tc{r}\tc{r}}-\varphi_{\tc{r}\tc{l}})} \\
 &  &   & 1
\end{matrix} \right)
\end{equation}
There is no decoherence and the final state of the two masses is pure. We expect a pure state of two qubits to have three independent phase factors (four components minus a total phase). Indeed, tven though we have four different phases in~\eqref{rhofBMVfull}, they only appear in three linearly independent combinations. For example, taking the sum/difference of the combinations $\varphi_{\tc{ll}}-\varphi_{\tc{lr}}$, $\varphi_{\tc{ll}}-\varphi_{\tc{rl}}$, $\varphi_{\tc{rr}}-\varphi_{\tc{lr}}$ we can recover all other phase factors that appear in~\eqref{rhofBMVfull}. To leading order in $T$ we have
\begin{align}
\label{eq:BMVphases}
\varphi_{\tc{ll}}&=\varphi_{\tc{rr}}= \frac{{\lambda}_{\tc{a}}{\lambda}_{\tc{b}} T}{4\pi\hbar D }+ \mathcal{O}(D/T), \nonumber \\ 
\varphi_{\tc{l}\tc{r}}&= \frac{{\lambda}_{\tc{a}}{\lambda}_{\tc{b}} T}{4\pi\hbar \left(D+\Delta x\right)}+ \mathcal{O}((D+\Delta x)/T), \nonumber \\
\varphi_{\tc{r}\tc{l}}&= \frac{{\lambda}_{\tc{a}}{\lambda}_{\tc{b}} T}{4\pi\hbar \left(D-\Delta x\right)}+ \mathcal{O}((D-\Delta x)/T)
\end{align}
\textcolor{red}{}
To make the analogy with the gravitational force, we may set the coupling constants to ${\lambda}_a\equiv m_a \sqrt{4\pi G\,}$. Then, ~\eqref{eq:BMVphases} correspond to the phases that appear in the Newtonian descriptions of the entanglement-through-gravity proposals. Note that in the entanglement-through-gravity protocols, typically there is an extra symmetry due to A and B having the same `splitting distance' $\Delta x$. This leads to a further condition $\varphi_{\tc{ll}}=\varphi_\tc{rr}$ and two linearly independent combinations of phases in the final state.



\end{document}